\documentclass{aastex61}

\newcommand\aastex{AAS\TeX}

\received{ }
\revised{ }
\accepted{\today}
\submitjournal{ApJ}

\shorttitle{\aastex\ INTRIGOSS: Synthetic Spectral Library}
\shortauthors{Franchini et al.}

\begin{document}

\title{Gaia-ESO Survey: INTRIGOSS - A new library of High Resolution Synthetic Spectra}

\correspondingauthor{Mariagrazia Franchini}
\email{mariagrazia.franchini@inaf.it}

\author[0000-0001-5611-2333]{Mariagrazia Franchini}
\affil{INAF - Osservatorio Astronomico di Trieste, Via G. B. Tiepolo 11, Trieste, I--34143, Italy }

\author{Carlo Morossi}
\affiliation{INAF - Osservatorio Astronomico di Trieste, Via G. B. Tiepolo 11, Trieste, I--34143, Italy }

\author{Paolo Di Marcantonio}
\affiliation{INAF - Osservatorio Astronomico di Trieste, Via G. B. Tiepolo 11, Trieste, I--34143, Italy }

\author{Miguel Chavez}
\affiliation{Instituto Nacional de Astrof\'isica, \'Optica y Electr\'onica, Luis Enrique Erro 1,  72840 Tonantzintla, Puebla, Mexico}

\author{Gerry Gilmore}   
\affiliation{Institute of Astronomy, University of Cambridge, Madingley Road, Cambridge CB3 0HA, United Kingdom}

\author{Sofia Randich}      
\affiliation{INAF - Osservatorio Astrofisico di Arcetri, Largo E. Fermi 5, Florence, I-50125 , Italy}

\author{Ettore Flaccomio}   
\affiliation{INAF - Osservatorio Astronomico di Palermo, Piazza del Parlamento 1, Palermo, I-90134, Italy}

\author{Sergey E. Koposov}  
\affiliation{Institute of Astronomy, University of Cambridge, Madingley Road, Cambridge CB3 0HA, United Kingdom}
\affiliation{McWilliams Center for Cosmology, Department of Physics, Carnegie Mellon University, 5000 Forbes Avenue, Pittsburgh, PA 15213, USA}

\author{Andreas J. Korn}    
\affiliation{Department of Physics and Astronomy, Uppsala University, Box 516, SE-751 20 Uppsala, Sweden}

\author{Amelia Bayo}    
\affiliation{Instituto de F\'isica y Astronomi\'ia, Fac. de Cencias, Universidad de Valparai\'iso, Gran Breta\~{n}a 1111, Playa Ancha, Chile}

\author{Giovanni Carraro}   
\affiliation{Dipartimento di Fisica e Astronomia, Universit\`a di Padova, Vicolo dell'Osservatorio 3,  Padova,   I-35122, Italy}

\author{Andy    Casey}      
\affiliation{Monash Centre for Astrophysics, School of Physics \& Astronomy, Monash University, Clayton 3800, Victoria, Australia}

\author{Elena Franciosini}  
\affiliation{INAF - Osservatorio Astrofisico di Arcetri, Largo E. Fermi 5,  Florence, I-50125, Italy}

\author{Anna    Hourihane}  
\affiliation{Institute of Astronomy, University of Cambridge, Madingley Road, Cambridge CB3 0HA, United Kingdom}

\author{Paula Jofr\'e  }    
\affiliation{Institute of Astronomy, University of Cambridge, Madingley Road, Cambridge CB3 0HA, United Kingdom}
\affiliation{N\'ucleo de Astronom\'{i}a, Facultad de Ingenier\'{i}a, Universidad Diego Portales, Av. Ej\'ercito 441, Santiago, Chile }

\author{Carmela Lardo}      
\affiliation{Laboratoire d'astrophysique, Ecole Polytechnique F\'ed\'erale de Lausanne (EPFL), Observatoire de Sauverny, CH-1290 Versoix, Switzerland}

\author{James Lewis }       
\affiliation{Institute of Astronomy, University of Cambridge, Madingley Road, Cambridge CB3 0HA, United Kingdom}

\author{Laura    Magrini}   
\affiliation{INAF - Osservatorio Astrofisico di Arcetri, Largo E. Fermi 5, Florence, I- 50125, Italy}

\author{Lorenzo Morbidelli} 
\affiliation{INAF - Osservatorio Astrofisico di Arcetri, Largo E. Fermi 5, Florence  I-50125,  Italy}

\author{G. G. Sacco}      
\affiliation{INAF - Osservatorio Astrofisico di Arcetri, Largo E. Fermi 5, Florence, I-50125, Italy}

\author{Clare Worley}    
\affiliation{Institute of Astronomy, University of Cambridge, Madingley Road, Cambridge CB3 0HA, United Kingdom}

\author{Tomaz Zwitter}   
\affiliation{Faculty of Mathematics and Physics, University of Ljubljana, Jadranska 19, 1000, Ljubljana, Slovenia}



\begin{abstract}

We present a  high resolution synthetic spectral library, INTRIGOSS, designed for studying FGK stars. The library  is based
on atmosphere models computed with specified individual element abundances via ATLAS12 code.
Normalized  SPectra (NSP) and surface Flux SPectra (FSP), in the 4830-5400\,\AA\, wavelength range, were computed  with the SPECTRUM code.
INTRIGOSS  uses  the  solar composition  by \citet{GRE07} and four  [$\alpha$/Fe] abundance ratios and consists 
of 15,232 spectra.
The synthetic spectra are computed with  astrophysical {\it gf}-values  derived by comparing synthetic predictions with a very high SNR solar spectrum 
and the UVES-U580 spectra of five cool giants. 
The validity of the NSPs is assessed by using the UVES-U580 spectra of 2212 stars observed in the framework of the Gaia-ESO Survey
 and characterized by  homogeneous and accurate  atmospheric parameter values and by detailed chemical compositions.
The greater accuracy of NSPs  with respect to spectra from the AMBRE, GES\_Grid, PHOENIX, C14,  and B17 synthetic spectral libraries is demonstrated
by evaluating the consistency  of the predictions of the different libraries for the UVES-U580 sample stars.
The validity of the FSPs is checked  by comparing their prediction with both observed spectral energy distribution and spectral indices.
The comparison of  FSPs  with  SEDs derived from   ELODIE, INDO--U.S.,  and MILES libraries indicates that the former reproduce the observed flux distributions
within a few percent and without any systematic trend. The good agreement between observational and synthetic Lick/SDSS indices
shows that the predicted blanketing of FSPs well reproduces the observed one, thus confirming the reliability of INTRIGOSS 
FSPs. 
\end{abstract}




\section{Introduction} \label{sec:intro}

The use of stellar spectral libraries dates back to several decades with some of the most prominent examples being the data set constructed by 
\citet{JACOBY84}  and that of \citet{KURUCZ79} which provide examples, respectively, of empirical and theoretical approaches to understand the stellar atmospheres, in particular the 
photospheres. Over the nearly 40 years since those papers a wealth of works have furnished with very extensive spectral libraries that 
have (partially) coped with the drawbacks found in the different approaches. On the empirical side, newer databases contain higher 
resolution spectra (e.g. ELODIE \citep{MO04}; Sloan Digital Sky Survey (SDSS) as, for example,  SEGUE \citep{YANNY09} and APOGEE 
\citep{MAJ17}; Galactic Archaeology with HERMES (GALAH) survey \citep{deSI15}; Gaia-ESO Survey  \citep[GES,][]{GIL12, RAN13}; etc.) 
with a much comprehensive coverage of the parameter space
(i.e. in effective temperature, $T_{\rm eff}$, surface gravity, log\,$g$,  and chemical composition, [Fe/H] and [$\alpha$/Fe]), and wavelength.
On the theoretical side, current available libraries of theoretical spectra have incorporated more extended line lists and updated atomic and 
molecular parameters in addition to a more detailed calculation of the radiative transfer equation allowing for departures of the local 
thermodynamic equilibrium and, in general, including more realistic treatment of the physical processes, e.g. motions, from microturbulence to 
3D dynamics of convection,  than those that characterize a classical model atmosphere \citep[][etc]{HUBENY88, HUSSER13}.

Stellar libraries have been  extensively applied in a number of astrophysical topics as: 
\begin{itemize}
 \item  automatic analysis of data in stellar surveys to derive  atmospheric parameters, radial and rotational velocities;
 \item detection of exoplanets via cross-correlation with spectra templates;
 \item study of star formation history of galaxies by using synthetic and observed photometric indices and/or Spectral Energy Distributions (SEDs).
\end{itemize}

It is practically impossible to review here the full set of applications, however, the compilation of \citet{LEITHERE96} gives evidence of 
the use of spectroscopic stellar data sets in the particular case of, perhaps, the most widely spread implementation: studies of galaxy structure and evolution.

It is worthwhile noticing that most of the empirical libraries carry on the imprints of the local properties
of the solar neighbourhood hampering the study of stellar populations characterized by star formation history different from that one in our vicinity.
Therefore, several theoretical libraries were computed to complement the empirical ones and are available in the literature
(e.g. \citealt{COELHO05}, C05; \citealt{deLA12}, AMBRE; \citealt{HUSSER13}, PHOENIX;  \citealt{COELHO14}, C14; \citealt{BRAHM17}, B17).
As mentioned before, these theoretical libraries can be used to derive stellar atmospheric parameters by comparing observed spectra with their predictions even if some limitations 
arise from the approximation and inaccuracies in the models and input data used to compute them.  

The computation of a theoretical library consists in the calculation of a set of model atmospheres giving the temperature, gas, electron and radiation pressure
distributions as a function of column mass or optical depth and the computation of the emerging spectra via a spectral synthesis code. As far as the set of 
atmosphere models the most commonly used for the analysis of Sun-like stars are those computed with ATLAS9 \citep{CA03}, ATLAS12  \citep{KU05a}, MARCS \citep{GU08}, or 
PHOENIX \citep{HAU99} codes, while synthetic spectra may be computed with several spectral synthesis codes like  DFSYNTHE \citep[][]{CA05, KU05a, KU05b}, 
SPECTRUM \citep{GR94}, PHOENIX, MOOG\footnote{\url{http://www.as.utexas.edu/~chris/moog.html}}, etc.

The available libraries in the literature differ not only because of the different model and synthesis codes used but also because of different adopted chemical mixture 
and different atomic and molecular line lists. Moreover, not all the libraries provide both normalized High Resolution (HR) spectra and SEDs 
since they require different approaches (see, for example, the discussion in C14). 

A comparison among  the performance of existing libraries is given in B17 where the authors presented their synthetic spectra library devoted to the 
determination of atmospheric stellar parameters via the Zonal Atmospheric Stellar Parameters Estimator (ZASPE). Even if the B17 spectral library is able to obtain 
results more consistent with the SWEET-Cat catalogue used to validate ZASPE code than C05 and PHOENIX, such a library, adopting solar scale abundances,
may introduce systematic biases in the determination of log\,{\it g} values for stars having non solar 
[Mg/Fe] (see B17). Furthermore, B17 library consists only of normalized HR spectra and therefore is not fully suited for stellar population studies
requiring SEDs. 

In this paper we present a new library of HR synthetic spectra, INaf-TRIeste Grid Of Synthetic Spectra  (INTRIGOSS), for F,G,K stars 
covering the wavelength range from 4830 to 5400\,\AA~which, even if rather narrow, is very useful to derive  stellar atmosphere parameter values
for these stellar types. INTRIGOSS spectral library,  which is available on the web\footnote{http://archives.ia2.inaf.it/intrigoss/}, aims to
furnish a tool  for stellar atmosphere parameter determination and it consists  of 
Normalized SPectra (NSP) and surface Flux SPectra (FSP).
Model atmospheres and theoretical spectra were obtained assuming as solar composition the one derived by \citet{GRE07}
and the full consistency of the library was guaranteed by using ATLAS12  and
SPECTRUM v2.76f  codes which allowed us to specify the same individual  element abundances both in 
deriving the atmosphere structures and the emerging spectra. 
The NSPs and FSPs were computed with an 
atomic and molecular line list built by tuning oscillator strengths in order to reproduce a set of HR reference spectra, namely 
the Solar spectrum and the GES spectra of  five cool giants with high Signal-to-Noise Ratios (SNR $>$ 100).  
The final line list includes also a sub-set of {\it bona fide} predicted lines.
The  predicted lines (PLs) are calculated absorption lines corresponding to transitions between levels  predicted by atomic structure codes but not 
measured in laboratory and are affected by large uncertainties in  their computed intensity and wavelength.
We call  {\it bona fide} PLs those with wavelength positions and 
oscillator strengths  consistent with the features in the HR reference spectra and which can be safely used to compute reliable SEDs.
The NSP and FSP spectra are computed  with a wavelength sampling of 0.01\,\AA{} thus allowing their degradation at any resolving power $R\lesssim$\,240,000. 

INTRIGOSS covers in the parameter space the following ranges:
$T_{\rm eff}$ 3750$\div$7000\,K,   log\,$g$  0.5$\div$5.0\,dex, and [Fe/H] -1.0$\div$+0.5\,dex.
Furthermore, to  also account for stars with non-solar scaled abundances of $\alpha$ elements,  four different chemical compositions
with [$\alpha$/Fe]=-0.25, 0.00, +0.25, and +0.50 were adopted.

The validity of the line list used in computing INTRIGOSS NSPs and FSPs and the improvement of  INTRIGOSS NSP spectra with respect to  those of the already available libraries 
is attested by comparing them  with a set of 2212 GES UVES spectra (hereafter UVES-U580 sample) obtained in a setup centered at 580nm (UVES-580) in the framework
of the iDR4 release of Gaia-ESO Survey. GES was designed to target all major Galactic components (i.e. bulge, thin and thick discs, halo and clusters), 
with the goal of constraining the chemical and dynamical evolution of the Milky Way  \citep{GIL12, RAN13}.
When completed, the survey will have observed with Fibre Large Array Multi Element Spectrograph/UV-Visual Echelle Spectrograph
(FLAMES/UVES) a sample of several thousand FGK-type stars within 2 kpc of the Sun in order to derive the detailed kinematic and
elemental abundance distribution functions of the solar neighbourhood.
The sample includes mainly thin and thick disc stars, of all ages and metallicities, but also a small fraction of local halo stars.
Data reduction of the FLAMES/UVES spectra has been performed using a workflow specifically developed for this project \citep{SA14}.
The iDR4 release contains radial and rotational velocities, recommended stellar atmosphere parameters, and individual element abundances. 
It also contains the stacked spectra derived from observations made for the Gaia-ESO Survey during 
iDR4 and tables of metadata summarizing these spectra. 
The UVES-580 spectra were analyzed with the Gaia-ESO multiple pipelines strategy, as described in \citet{SM14}. The results of each pipeline are combined 
with an updated methodology (Casey et al., in prep.) to define the final set of recommended values of the atmospheric parameters and chemical 
abundances that are part of iDR4 \citep[see also][]{MA17}.

The validity of the INTRIGOSS FSP spectra is checked by comparing observed and synthetic SEDs for a sub-sample of stars and through the analysis of 
synthetic photometric indices computed for all the UVES-U580 stars from the corresponding derived FSPs. In particular, in this paper, we show 
the very good agreement  between  observational and  INTRIGOSS synthetic 
  Mg$_1$, Mg$_2$, and Mgb  Lick/SDSS indices  \citep{FR10}.

In Section \ref{sec:library} we describe how the models and the synthetic spectra were computed,  the atomic and molecular line lists we used, and 
the sample of UVES-U580 spectra adopted as reference to check the reliability of our library. 
Section\,\ref{sec:comparison} compares the predictions of INTRIGOSS with those of publicly available spectral libraries and
discusses the achieved improvements.  In Section\,\ref{sec:data} we describe 
the INTRIGOSS data products and auxiliary files available on the web. 
Finally, in  Section\,\ref{sec:summary} we summarize and conclude.

\section{The theoretical library of synthetic spectra: INTRIGOSS} \label{sec:library}

The construction of a theoretical stellar library requires several fundamental ingredients
that are applied along different steps. The two main ingredients are: (a) a set of model 
atmospheres that are needed as input in the calculation of synthetic spectra and (b) a fiducial 
line list to take into account the relevant individual atomic  as well as
 molecular transitions expected to be important for the parameter space under consideration.
Additionally, one requires the appropriate codes for the computations and the choice of 
the spectrum/spectra to which the theoretical data set will be compared to check the 
capability of the library in representing real stars.

The construction of INTRIGOSS consists of several steps. The first is the calculation of a 
grid of model atmospheres that provide the variation of physical quantities throughout the atmosphere. This task has been 
carried out by using the ATLAS12 code developed by \citet{KU05a}. Once a set of theoretical 
models has been constructed, the next step is to calculate, under the same considerations (i.e.
Local Thermodynamic Equilibrium, LTE),  the corresponding synthetic spectra at high resolution. 
If the  input fiducial atomic and molecular line list 
is adequate to predict line profiles of observed spectra, this would be the second 
and last step in the process, nevertheless, any working line list actually needs fine 
tuning of the main agent that affect line intensities, i.e. the oscillator strengths or the log({\it gf})
values, for the (at least) more prominent transitions. This last task is usually conducted by 
comparing synthetic spectra with observed data of the highest quality. Below we provide details of 
these processes that end-up with the theoretical library INTRIGOSS.

\subsection{The atmosphere models} \label{subsec:models}
As discussed in \citet{KU05a} and \citet{CA05}, ATLAS12 can generate an atmospheric model for any 
desired individual element chemical composition and microturbulent velocity ($\xi$), since the treatment 
of opacity is based on the Opacity Sampling technique, instead of the Opacity Distribution Functions 
method used, for example, by ATLAS9. We adopted for the reference solar abundances those obtained from 
\citet{GRE07} which have a wide consensus in the literature and whose validity is also confirmed, 
within the quoted uncertainties, by the abundance determinations derived by the Working Group 11 
(WG11) of the Gaia-ESO consortium \citep{MA17} from the analysis of  solar and M67 giant UVES 
spectra obtained with the U580 and U520 setups \citep{DEK00}.

We computed sets of atmosphere models in the following ranges of atmospheric parameters:
\begin{enumerate}
 \item $T_{\rm eff}$  from 3750 to 7000\,K at a step of 250\,K;
 \item log\,$g$ from 0.5 to 5.0\,dex at a step of 0.5\,dex;
 \item $[$Fe/H$]$ from -1.0 to +0.5\,dex at a step of 0.25\,dex;
 \item $[\alpha$/Fe$]$ from -0.25 to +0.5\,dex at a step of 0.25\,dex where the $\alpha$ elements 
considered are O, Ne, Mg, Si, S, Ar, Ca, and Ti;
 \item $\xi$ equal to 1 and 2\,km\,s$^{-1}$.
\end{enumerate}

Our atmosphere models were calculated by using ATLAS12 and, as initial starting models, the ATLAS9 Model 
Atmospheres calculated for the APOGEE survey available online\footnote{http://www.iac.es/proyecto/ATLAS-APOGEE/}.    
The resulting models computed with the two $\xi$ values, were checked by looking at the behaviors  
of temperature, gas pressure, electron density, Rosseland absorption coefficient, and radiation 
pressure  at all Rosseland optical  depths. The few gaps in the coverage of the resulting atmosphere 
models in the   $T_{\rm eff}$-log\,$g$ plane correspond to the absence of initial models in the  
APOGEE set.  In running ATLAS12 we used in input  atomic and molecular species 
(file {\it molecules.dat}\footnote{http://kurucz.harvard.edu/programs/atlas12/}), the lines of all the elements 
in the first 5 stages of ionization ({\it fclowlines.bin}\footnote{http://kurucz.harvard.edu/linelists/linescd/ \label{footnote lines}}),  
the  lines of diatomic molecules ({\it diatomicsiwl.bin}\textsuperscript{\ref{footnote lines}} and
{\it tioschwenke.bin}\footnote{http://kurucz.harvard.edu/molecules/tio/} by \citealt{SC98}), and the lines of
H$_2$O ({\it h2ofastfix.bin}\footnote{http://kurucz.harvard.edu/molecules/h2o/}) by \citet{PA97}.
The adopted steps along $T_{\rm eff}$, log\,$g$, $[$Fe/H$]$, and $[\alpha$/Fe$]$ correspond, roughly, to twice 
the standard uncertainties in the atmosphere parameter determinations quoted in the literature (see for 
example \citealt{MA17}). It is worth noting, that on the basis of the considerations presented in 
Sections\,\ref{subsec:NSPs} and \ref{subsec:FSPs}, a smaller step in temperature would make safer
any interpolation  required for  deriving stellar atmospheric parameters.
Therefore, in a future extended version of INTRIGOSS aimed to increase, in particular, its wavelength coverage,
we plan to adopt a finer  temperature  grid.

\subsection{The synthetic spectra} \label{subsec:spectra}
To obtain from each atmosphere model the corresponding emergent flux  and normalized spectrum  we used  
SPECTRUM\,v2.76f. SPECTRUM is a stellar spectral synthesis program that calculates, under the 
LTE approximation, the synthetic spectrum for a given model atmosphere. 
The code additionally requires line lists of atomic and molecular transitions, that should be as accurate 
and complete as possible, and supports all expected atomic elements and the following diatomic molecules: 
H2,  CH,  NH, OH, MgH, SiH, CaH, SiO, C2, CN, CO and TiO. The code can deliver  both the stellar-disk-integrated 
normalized spectrum and the absolute monochromatic flux at the stellar surface. 
The user should specify the wavelength range and sampling,  the value of microturbulence 
velocity ($\xi$), and the individual element abundances to be used. For this work we have built a line list 
by merging the line data used by \citet[][hereafter LO11]{LO11} and the cool5.iso.lst line list kindly provided 
to us by R. O. Gray (2011, private communication), complemented with molecular lines of CH, NH, MgH, SiH, 
C$_2$, CN, and TiO and with both atomic (in the three main expected stages of ionization) and molecular predicted 
lines from Kurucz's site.  The lines included in our calculations for the wavelength interval 4830 to 5400\,\AA{} 
are 1427628 including 16531, 339652, and  1071445 transitions for atomic, molecular, and PL entries, respectively. 

As indicated previously, accurate astrophysical synthetic spectra computation requires reliable atomic 
and molecular data, in particular accurate oscillator strength, {\it gf}-values,
for the transitions expected in the wavelength interval of interest. In recent years a number of online databases 
(i.e. NIST\footnote{https://physics.nist.gov/PhysRefData/ASD/lines\_form.html},
VALD\footnote{http://vald.astro.univie.ac.at/$\sim$vald3/php/vald.php},
NORAD\footnote{http://www.pa.uky.edu/$\sim$peter/newpage/}, the
Kurucz website, etc) provide line data  from a large variety of sources in the scientific literature. 
The {\it gf}-values given in the databases may either have been determined in the laboratory or derived 
from theoretical calculations. Thus the accuracy of these {\it gf}-values may vary widely from line to line; 
some are known with accuracies of 1\% or better, while others may be off by  orders of magnitude.
A possible way to reduce these uncertainties is to compare high SNR  spectra of stars, with well known 
atmospheric parameters and abundances, with their corresponding computed synthetic spectra. In this way the 
{\it gf}-values may be checked (and if needed,  adapted  with a trial-and-error strategy) by looking 
for a best agreement between the synthetic and   observed line profiles thus deriving astrophysical 
{\it gf}-values. Since the line profiles depend both on stellar characteristics, namely $T_{\rm eff}$, 
log\,$g$, element abundance, and $\xi$, and on the {\it gf}-value, the risk in such an approach is 
to wrongly adapt the {\it gf}-values to compensate for potential inaccuracies in the assumed
values of  atmospheric parameters and in the modeling assumptions. It is, therefore,
important to perform the comparison of the synthetic and observed profile of the same line in spectra of 
as many (and as different) as possible stars in order to disentangle the effect of incorrect {\it gf}-values 
from those due to uncertainties in  the other parameters. 

In this context, \citet[][hereafter LO11]{LO11} used the high-resolution spectrum of three main sequence stars, 
the Sun, Procyon, and $\epsilon$\,Eri, characterized by the following $T_{\rm eff}$, log\,$g$, and $\xi$
values: SUN\,(5777,4.438,1.1), Procyon\,(6550,4.0,1.2), and  $\epsilon$\,Eri\,(5050,4.5,0.55) and  
assuming solar composition from \citet{AND89} also for the last two stars.
He used the solar spectrum observed in 1981 with the NSO/KPNO Fourier Transform
Spectrograph (FTS), degraded at  R$\sim$80,000,  and,  for Procyon and $\epsilon$\,Eri, 
several optical spectra taken  with the Hermes spectrometer
on the 1.2 m Mercator telescope at La Palma Observatory, Canary Islands. The comparison 
synthetic spectra were calculated with the LTE radiative transfer 
SCANSPEC\footnote{alobel.freeshell.org/scan.html} code and, in such a way,  LO11 updated 
the log({\it gf}) values of 911 neutral lines in the wavelength range 4000$\div$6800\,\AA{}.
The main causes of uncertainties in the LO11 results arise from: i) the problems of deriving 
the solar intensity (averaged over the solar disk) from the NSO/KPNO Fourier Transform
observations; ii) the assumption of solar composition for Procyon, and $\epsilon$\,Eri even 
if some differences in individual element abundances are reported in the literature
\citep[see for example][]{JO15}; iii) the use of only relatively high temperature ($T_{\rm eff}>$ 5000\,K)  
main sequence stars which does not allow to check the log({\it gf}) 
values of those atomic and molecular lines that are mainly prominent in giants and/or cooler stars.

We decided to complement the LO11 work by performing the same kind of  analysis but by using an 
{\it ad hoc} derived high SNR solar spectrum and the UVES-U580 spectra 
with SNR above 100 of five giant stars (see Table\,\ref{tab:giants}) with atmospheric parameters
in the following ranges: $T_{\rm eff}$ between 4500 and 5000\,K, log\,$g$ from 2.0 to 3.2\,dex, and  
$\xi$ from 1.0 to 1.5\,km\,s$^{-1}$, and, for each star, the individual
element abundances derived by GES Consortium and reported in the GESiDR4Final catalogue \citep{MA17}.

The outline of our method is the following:
\begin{itemize}
 \item use of the solar spectrum as the main reference to derive the astrophysical log({\it gf}) values for atomic and 
 molecular lines that are important at solar effective temperature and 
 gravity by assuming no uncertainties in the solar parameters and in the adopted atmosphere model;
 \item use of the five giant spectra to derive the astrophysical log({\it gf}) values of those  
 lines that are more prominent at temperatures and gravities lower than the
  solar ones and to fix globally the MgH opacity by using the scaling  factor {\it f$_{\rm MgH}$} (see Section\,\ref{subsubsec:giants});
 \item use of a large sample of stars (more than 2200) covering a wider range of atmospheric 
 parameter values to validate the final list of astrophysical log({\it gf})  values. 
\end{itemize}

\subsubsection{Refinement of oscillator strengths {\bf and tuning of the central wavelengths}: The solar case}
\label{subsubsec:sun}
We used an observed solar spectrum which is the average of 59 integrated sunlight spectra, 
as reflected by the Moon, taken with HARPS spectrograph at the 3.6-m La Silla European Southern
Observatory (ESO) telescope. These spectra are the out-of-transit sub-sample  of those   taken 
to detect  the Rossiter-McLaughlin effect in the Sun due to the Venus transit of 2012 June 6 \citep{MO13}. 
The SNR of the average spectrum, as evaluated by looking at the  ratio between the mean flux
and the standard deviation of the mean at each wavelength, is about\,4000. Then the spectrum was 
degraded at the resolution of the Hermes spectra (R=80,000).

The normalized synthetic solar spectrum to be used in the comparison was computed with SPECTRUM 
starting from the ATLAS12 model obtained with $T_{\rm eff}$=5777\,K, log\,$g$=4.4377\,dex, 
$\xi$=1.0\,km\,s$^{-1}$, and with the solar abundances by \cite{GRE07}. This spectrum was 
then degraded at R=80,000 and convolved with the geometric mean of the solar  $v\sin i$ and 
macroturbulence velocity values (2.5\,km\,s$^{-1}$). 

Prior to conducting modifications on the line data we first need to normalize the observed 
spectra, a crucial process to derive reliable log({\it gf})s. With the goal of matching the 
continuum levels of observed and synthetic spectra, we searched 
for continuum flux reference points in the normalized synthetic spectrum and  identified 
the wavelength intervals with flux levels in excess of 0.99, avoiding, in this way, regions with absorption 
lines where the uncertainties in the log({\it gf}) values may play a role.  
Then, the observed spectrum in each of the corresponding wavelength regions is divided by the synthetic 
one and these ratios are fitted with a polynomial. Eventually, the observed spectrum is divided by 
the so computed polynomial to obtain the normalized spectrum.

A trial-and-error procedure based on the comparison between the normalized and the synthetic  
solar spectrum is now applied to modify (when needed) the input log({\it gf}) values 
{\bf and/or the central wavelengths} in order 
to match the profile  of the observed lines or blends.  We compared our observed and synthetic 
solar spectra to adapt the log({\it gf})s of those lines 
which are responsible of clearly detectable flux minima in the observed and/or synthetic spectra to 
obtain their astrophysical values. 
We followed a three steps approach:
\begin{enumerate}
\item  we looked in our input line list to identify
the transitions responsible for those minima present in both the observed and synthetic spectra
and we iteratively modified their log({\it gf}) values until we found a satisfactory agreement.
In general we stopped the iteration when the difference between synthetic and observed spectra 
were below $\pm$ 0.003;
 \item   we selected the minima present only in the synthetic spectra
and reduced the log({\it gf}) values of the corresponding theoretical lines to match the
observed spectrum;
 \item   we searched in our input line list and in the  online databases listed in 
Section\,\ref{subsec:spectra} for theoretical lines
which may correspond to minima present only in the observed spectrum and,  for those lines found, we fine tuned their  log({\it gf}) values
and added them to our input list if needed.
\end{enumerate}

\noindent {\bf In all the above listed steps we also checked the central wavelength of each identified features and, in a few cases, we slightly modified it
to better match the observed spectrum.
}
Eventually, the identification of the lines corresponding to the minima was double-checked by looking at the same wavelength regions in the synthetic
and observed spectra of five giants (see Section\,\ref{subsubsec:giants}). 

In conclusion we derived astrophysical log({\it gf}) values
for 2229 lines, that include 850, 35, and 1344 atomic, molecular, and predicted transitions, respectively. The 850 atomic lines
include 100 lines from LO11 that required slight modification of their log({\it gf})s 
or of their central wavelength to get a better match of the spectra. These corrections are mainly 
due to the slightly different solar abundances adopted, by the higher SNR of our solar spectrum and by 
the inclusion in our synthetic spectrum of molecular and predicted lines non present in 
the SCANSPEC spectra of LO11.

\subsubsection{Refinement of oscillator strengths: five giant stars}
\label{subsubsec:giants}
The above procedure should, in principle, be sufficient to calculate fiducial theoretical spectra 
that represent stars with atmospheric parameters close to solar. However, to account for potential 
targets of lower temperature and surface gravity one needs to extend the log({\it gf}) tuning analysis 
to giant stars. For this extension we considered the five giants in Table\,\ref{tab:giants}.

\begin{figure}[ht!]
\includegraphics[width=\textwidth]{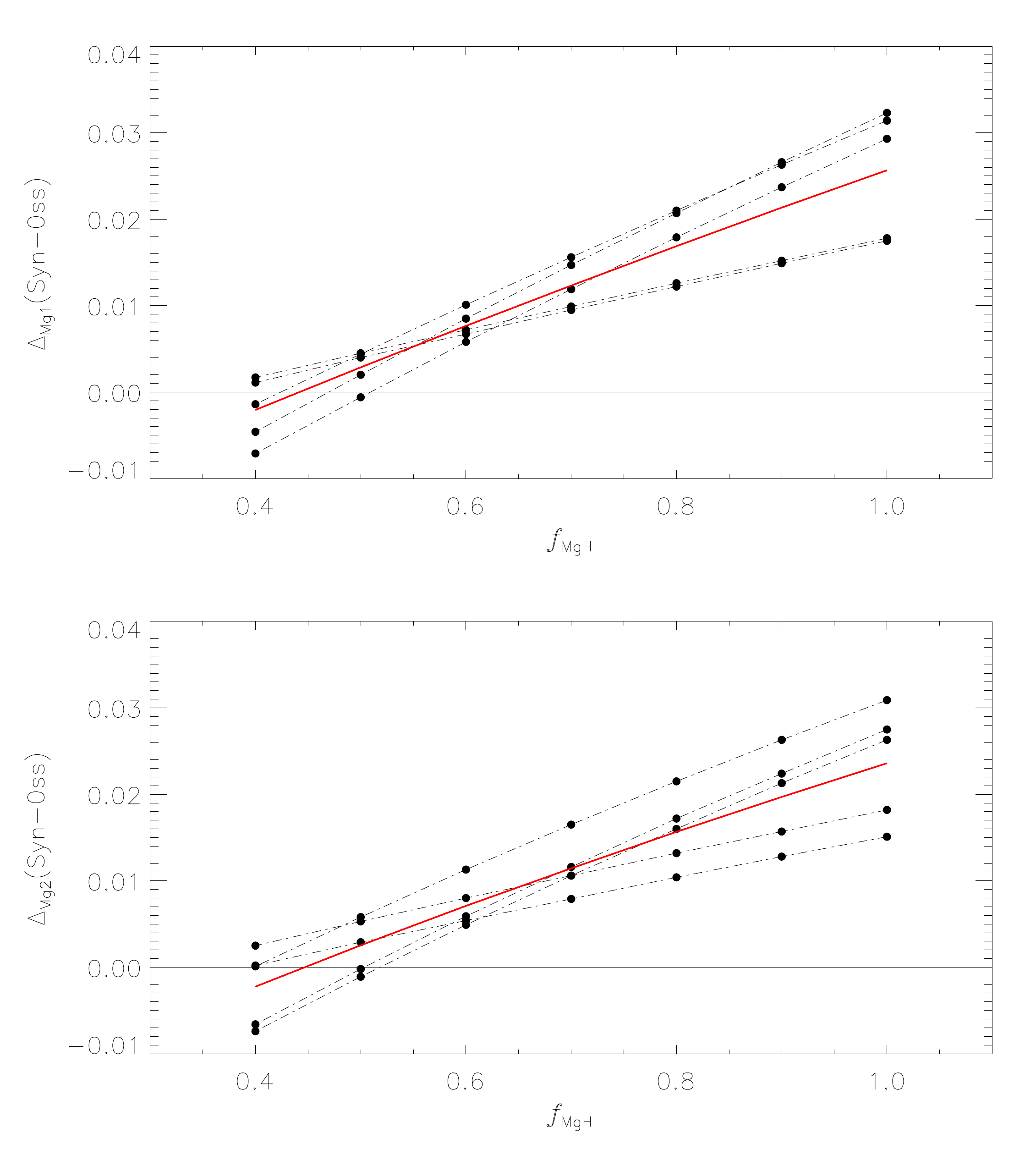}
\caption{Differences between synthetic and observed Mg$_1$ and Mg$_2$ Lick/SDSS indices for the five giant stars in Table\,\ref{tab:giants} (black dot-dashed thin lines)
and  their average values (red thick line) as a function of the $f_{\rm MgH}$ value used in computing the synthetic spectra (see text).\label{fig:indgiant}}
\end{figure}

We computed for each i-th star its synthetic spectrum (S$_{\rm i}$) by using the GES atmospheric 
parameter values, the individual element abundances and the  line list which includes the astrophysical  log({\it gf})s
 derived  from the solar spectrum  analysis.
 Then, we adopted the  same procedures used for 
normalizing the observed solar spectrum for each i-th giant to derive from its UVES-580 spectrum the 
normalized one (O$_{\rm i}$). First of all we checked that the modified  log({\it gf}) values obtained in 
Section\,\ref{subsubsec:sun} provide a good agreement of synthetic and observed spectra also for these five stars.
We adopted as acceptance threshold a value of $\pm 0.01$ which is larger than that one used for the Sun due to 
the lower resolution (R=47,700) and SNR ($\sim 100$). Actually,  in all but a very few cases, we did not need to 
go back to the analysis of the solar spectra to re-tune the log({\it gf})s.
Then, we looked for features which were present only in the spectra of the giants. 
By adopting the same trial-and-error strategy used for the solar case but the new acceptance threshold we were able to derive
astrophysical  log({\it gf}) values for an additional number of 175 lines, namely 49, 42, and 84 atomic, molecular,
and predicted transitions, respectively. As far as the large number of weak lines of MgH, which is the dominant molecular opacity contributor 
in our wavelength range, are concerned  we decided to check individually, for this molecule, only the 
log({\it gf})s of the strongest features. The contribution of the other  MgH lines was then 
fine-tuned by means of a scaling factor, {\it f$_{\rm MgH}$}, by which their  
{\it gf}s are multiplied (see SPECTRUM documentation). To evaluate the most appropriate 
value of  {\it f$_{\rm MgH}$} we computed the Lick/SDSS Mg$_1$ and Mg$_2$ indices \citep{FR10}, which 
are strongly affected by the MgH lines, from the UVES-U580 spectra and compare them with 
those from synthetic spectra calculated with {\it f$_{\rm MgH}$} in the range 
1.0$\div$0.4. Figure\,\ref{fig:indgiant} shows that, on the average, the best agreement 
is obtained with  {\it f$_{\rm MgH}$}=0.45.

It is important to remark that, should we consider that the need of such a low  {\it f$_{\rm MgH}$} 
is ascribed to uncertainties in the Mg abundances, one would require to decrease the Mg abundances, 
log(Mg/H), of  all the five giants  by $\sim$0.35\,dex  in order to keep the {\it f$_{\rm MgH}$} value at 1.0.
On the other hand, such a low Mg abundances are inconsistent with those derived from  
the analysis of atomic Mg lines. Therefore, we are confident that the obtained {\it f$_{\rm MgH}$}=0.45  
is not due to a wrong GES Magnesium abundance determination but to an overestimate of the MgH 
opacities  computed by \citet{KUR14} as also found and discussed by \citet{WE03}. Thus, hereafter we adopted 
the value {\it f$_{\rm MgH}$}=0.45 to correct such overestimate and to be consistent with the 
log(Mg/H) derived from atomic lines.

\begin{deluxetable*}{cccrrrrrrrrrrc}[t!]
\tablecaption{Giant stars used to derive astrophysical log({\it gf}) values \label{tab:giants}}
\tablecolumns{14}
\tablewidth{0pt}
\tablehead{
\colhead{Cname} & \colhead{ $T_{\rm eff}$} & \colhead{$\sigma_{T_{\rm eff}}$} & \colhead{log\,$g$} & \colhead{$\sigma_{{\rm log}\,g}$} & \colhead{[Fe/H]} & 
\colhead{$\sigma_{\rm [Fe/H]}$} & \colhead{$\xi$} & \colhead{$\sigma_{\xi}$} & \colhead{$V_{\rm rad}$} & \colhead{$\sigma_{V_{\rm rad}}$} & \colhead{$v\sin i$} & 
\colhead{$\sigma_{v\sin i}$} & \colhead{SNR\tablenotemark{a}} \\
\colhead{ } & \colhead{K} & \colhead{K} &  \colhead{dex} & \colhead{dex} &  \colhead{dex} &  \colhead{dex} &  \colhead{km\,s$^{-1}$} & \colhead{km\,s$^{-1}$} &  
\colhead{km\,s$^{-1}$} &  \colhead{km\,s$^{-1}$} & \colhead{km\,s$^{-1}$} &  \colhead{km\,s$^{-1}$} & \colhead{}
}
\startdata
      00241708-7206106 &  4510 &   117 &   2.10 &   0.23 &  -0.70 &   0.10 &   1.34 &   0.08 &   2.91 &   0.57 &   2.15 &   2.82 & 108 \\
      00251219-7208053 &  4513 &   114 &   2.04 &   0.23 &  -0.67 &   0.10 &   1.45 &   0.03 & -34.30 &   0.57 &   2.12 &   2.61 & 105 \\
      00240054-7208550 &  4541 &   121 &   2.06 &   0.23 &  -0.71 &   0.09 &   1.40 &   0.07 & -13.13 &   0.57 &   2.10 &   2.56 & 108 \\
      02561410-0029286 &  4834 &   117 &   2.75 &   0.21 &  -0.70 &   0.09 &   1.05 &   0.05 & -68.65 &   0.57 &   2.39 &   2.85 & 113 \\
      03173493-0022132 &  4966 &   121 &   3.14 &   0.23 &  -0.63 &   0.10 &   1.03 &   0.07 & -40.28 &   0.57 &   2.22 &   2.51 & 113 \\
\enddata
\tablenotetext{a}{From the UVES-U580 FITS file headers.}
\tablecomments{Atmospheric parameter values from the recommended parameters and abundances table in the GESiDR4Final catalogue}
\end{deluxetable*}

\subsection{Assessment of the quality of the modified line list}

\begin{figure*}[ht!]
\includegraphics[width=\textheight,angle=90]{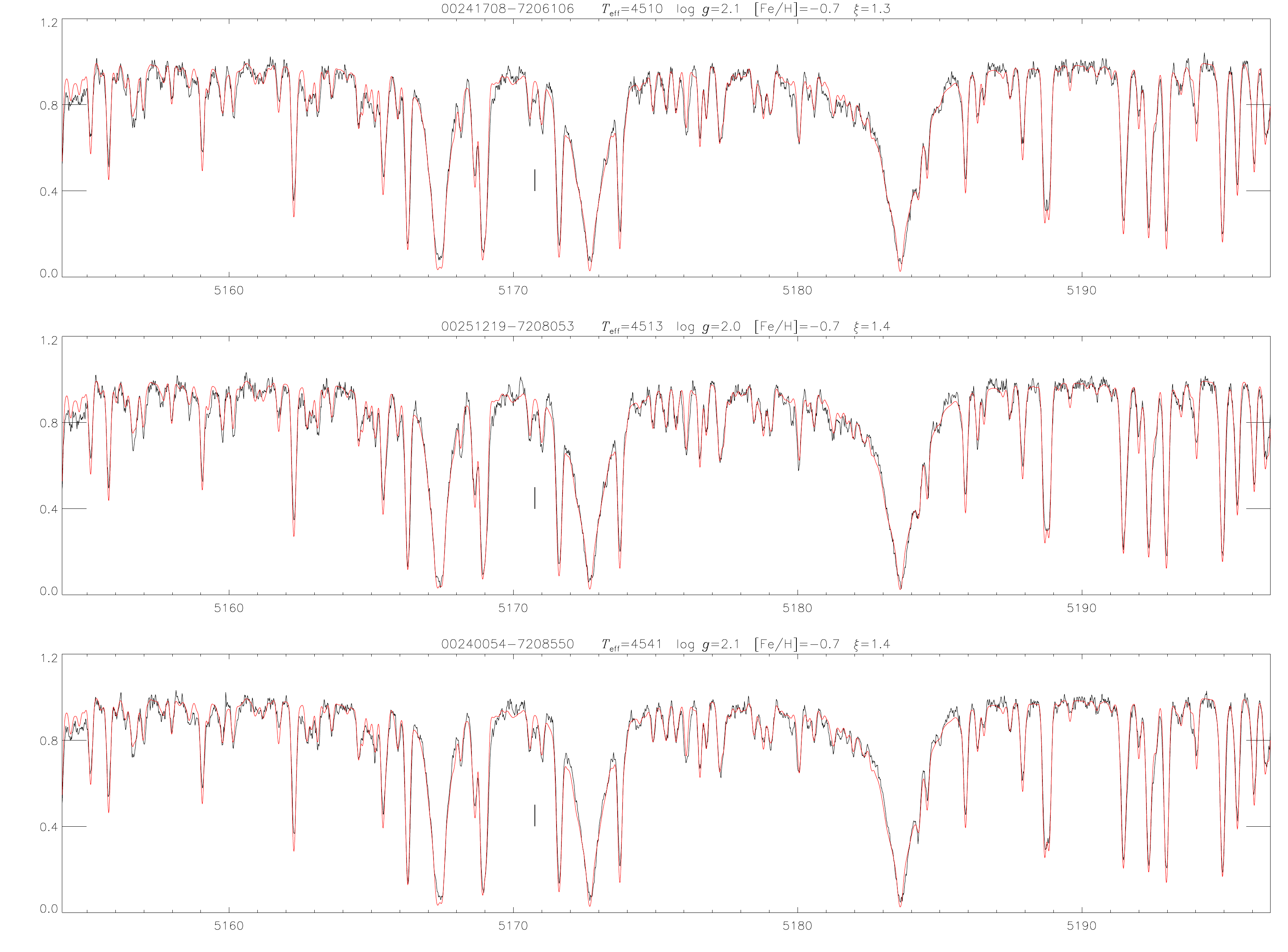}
\caption{Example of the comparison between the observed spectra (black lines) and the synthetic ones (red lines) computed by using the derived astrophysical log({\it gf}) 
for the first 3 giant stars in Table\,\ref{tab:giants}. As an example of unidentified lines the feature at 5170.77 is indicated (vertical bar).
\label{fig:gfexample1}}
\end{figure*}

\begin{figure*}[ht!]
\includegraphics[width=\textheight,angle=90]{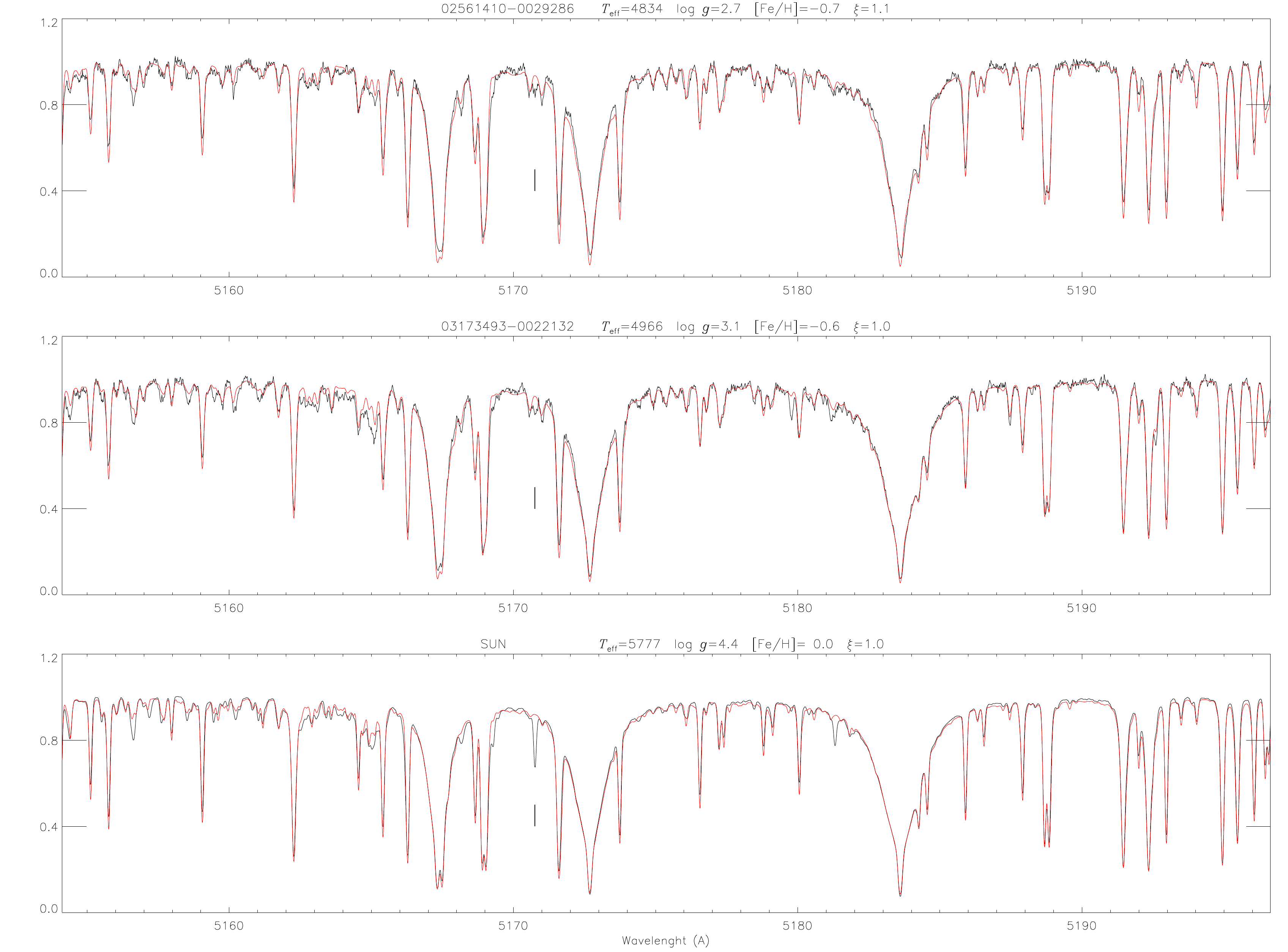}
\caption{Example of the comparison between the observed spectra (black lines) and the synthetic ones (red lines) computed by using the derived astrophysical log({\it gf}) 
for the last 2 giant stars in Table\,\ref{tab:giants} and for the Sun. As an example of unidentified lines the feature at 5170.77 is indicated (vertical bar).
\label{fig:gfexample2}}
\end{figure*}

Figures\,\ref{fig:gfexample1} and \ref{fig:gfexample2}  show an example of the  agreement between 
observed and synthetic spectra achieved by using the above-derived astrophysical log({\it gf})s. 
For the solar case, in which the uncertainties of the observed spectrum are negligible, a quantitative 
estimate of the improvement with respect to the use of the initial log({\it gf}) list is given by the 
decrease of the standard deviation of the residuals from $\sim 0.06$ to $\sim 0.03$. In  the case of 
the five giants, where  observational uncertainties must be taken into account, we calculated, 
as a figure of merit, $r^{\rm i}_{med}=median[(\frac{{\rm O_i}(\lambda)-{\rm S_i}(\lambda)}{\Delta {\rm O_i}(\lambda)})^2]$, 
where $\Delta {\rm O_i}(\lambda)$ were obtained from the inverse variance-per-pixel given in 
the UVES-U580 FITS files. In computing $r^{\rm i}_{med}$ we excluded those wavelength regions 
which were used to normalize the  O$_{\rm i}$ because they are bound to near-zero
residuals and are not sensitive to the quality of the used log({\it gf})s. We decided to use 
the median of the normalized residuals instead of the mean because the latter is strongly 
affected by the residuals in the region of unidentified lines. 
 As can be seen in Table\,\ref{tab:gigR} the use of the astrophysical log({\it gf})s allows 
 us to achieve a 30\% decrease of the $r_{med}$ values with respect to the initial ones.

\begin{deluxetable}{cccc}[ht!]
\tablecaption{Comparison of $r_{med}$ values for the five giants in Table\,\ref{tab:giants} 
when using  initial or astrophysical 
 log({\it gf})s \label{tab:gigR}}
\tablecolumns{4}
\tablewidth{0pt}
\tablehead{
\colhead{Cname} & \colhead{$r_{med}$ } & \colhead{$r_{med}$ } & \colhead{ratio} \\
\colhead{  } & \colhead{astrophysical} & \colhead{initial} & \colhead{\%} \\
}
\startdata
 00240054-7208550 &   1.46 &   2.10 &    69   \\
 00241708-7206106 &   1.30 &   1.84 &    71   \\
 02561410-0029286 &   1.33 &   1.96 &    68   \\
 00251219-7208053 &   0.89 &   1.37 &    65   \\
 03173493-0022132 &   0.76 &   1.14 &    67   \\
\enddata
\end{deluxetable}

We want to point out that some discrepancies  still  persist, in limited narrow wavelength 
regions, due to the presence in the observed spectra of  lines that we were not able to 
find in any of the atomic and molecular databases available in literature and, thus, 
being unidentified cannot be present in the synthetic spectra. A clear example of this situation 
is the feature at 5170.77\,\AA~ that is present in all observed spectra (see Figures\,\ref{fig:gfexample1}, \ref{fig:gfexample2}
and Table\,\ref{tab:lines}), 
and particularly prominent in the solar one.  In Appendix\,\ref{app:unident} we list the unidentified observed features.

\subsection{Validation of line list improvements on the GES sample}
\begin{figure}[h!]
\includegraphics[height=\textheight]{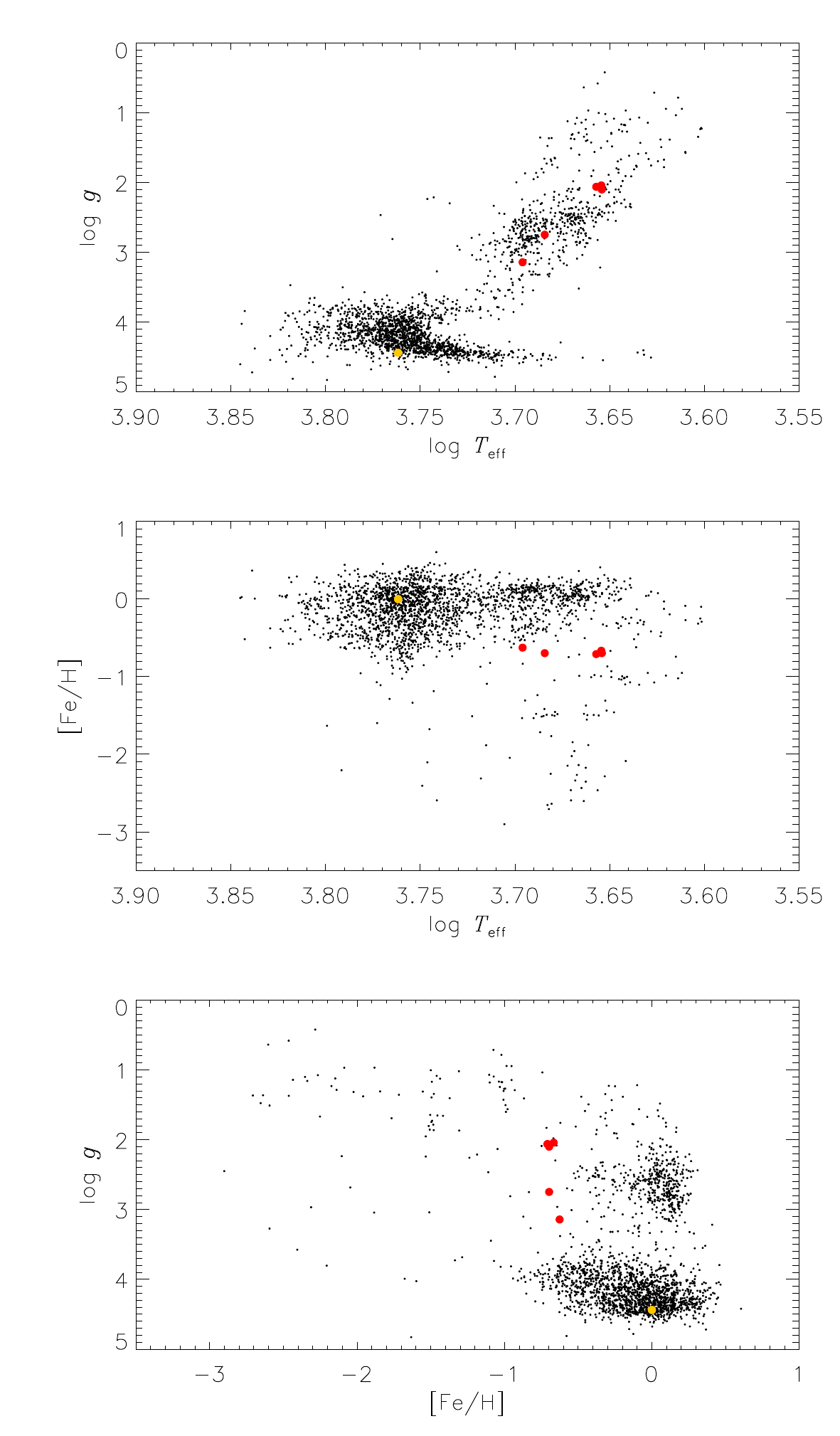}
\caption{GES atmospheric parameters of the 2212 stars in the UVES-U580 sample (black points) with superimposed those of the Sun (yellow circle) and of the five giants in Table\,\ref{tab:giants}
(red circles).
\label{fig:param}}
\end{figure}

To check the validity of our line list and, as a consequence, of the synthesized spectra over 
the full atmospheric parameter space covered by F,G, and K stars we decided to use a large
sample of stars with well known atmospheric parameter values and individual element abundances.
On what follows, we will use the observed spectra of  2212 UVES-U580 stars extracted 
from the forth Gaia-ESO (iDR4) release, whose atmospheric parameter coverage is shown in Figure\,\ref{fig:param},
in order to compare them  with the corresponding individual NSP synthetic spectra. Our check is based on the important
remark that, if we had derived wrong {\it gf}-values from the spectra 
of the Sun and of the five stars in Table\,\ref{tab:giants} because of errors in the adopted 
stellar atmosphere parameters, these {\it gf}-values should be very ineffective in reproducing the observed spectra of stars  
covering a much more extended parameter space. Furthermore, we have to point out that any coupling between 
our astrophysical {\it gf}-values and GES atmospheric parameter determinations is very unlikely since 
our main reference star in deriving astrophysical {\it gf}-values is the Sun (whose adopted atmospheric 
parameter values were not from GES). Moreover, the GES atmospheric parameter values of the five giants 
in Table\,\ref{tab:giants} are the homogenized results of several Working Group and Nodes of the 
Gaia-ESO consortium and are not at all related to  the process conducted in this work for calculating
theoretical models and  spectra.

\begin{figure}[h!]
\includegraphics[width=\textwidth]{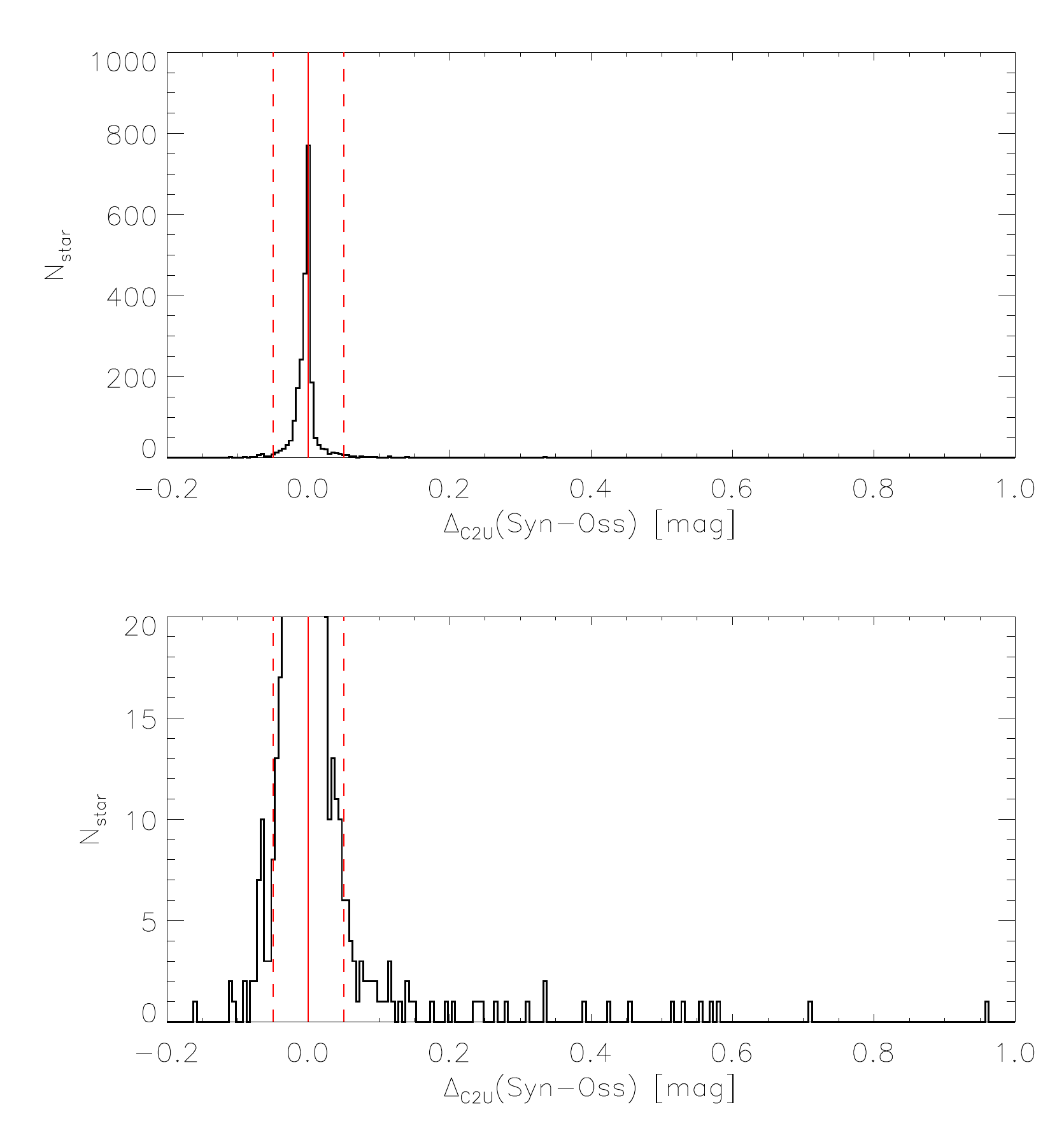}
\caption{Distribution of the differences between C2U indices derived for each pair of synthetic and  UVES-U580 spectra; the lower panel shows, by zooming the y-scale,
the presence of several outliers. Stars with $\Delta_{\rm C2U}$ values outside the red dashed lines ($\pm 0.05$\,mag) are removed from the UVES-U580 sample (see text).
\label{fig:C2U}}
\end{figure}

A first sample of 2616 stars was obtained by performing an SQL search to select all the
stars in the 3500-7000\,K and 0.25-5.25\,dex  effective temperature and surface gravity ranges
observed with U580 setup and characterized  by a Signal to Noise Ratio (SNR) greater than 10. 
Then,  we removed all the stars with some peculiarity flag and/or lack 
of an error estimate of the stellar atmosphere parameters. In such a way we obtained
a sample of 2311 stars well suitable for our analysis since it contains objects with homogeneously determined  
$T_{\rm eff}$, log\,$g$, detailed chemical composition, $\xi$, and  $v\sin i$ spanning the following ranges: 
$T_{\rm eff}$  from 3900 to 7000\,K; log\,$g$ from 0.4 to 4.9\,dex; $[$Fe/H$]$ from -2.9 to +0.6\,dex; 
$[\alpha$/Fe$]$ from -0.1 to +0.6\,dex; and $\xi$=0.1--3.0\,km\,s$^{-1}$. For each i-th star we run 
ATLAS12 and SPECTRUM codes, using its GES atmospheric parameter values, individual element 
abundances (for those elements with no estimate of [X/Fe] we assumed [X/Fe]=0) and our modified 
line list, to compute the appropriate normalized synthetic spectrum (S$_{\rm i}$) which is 
then used to obtain from the corresponding observed (stacked) UVES-U580 spectrum  
a normalized one (O$_{\rm i}$). The normalization was performed by applying the same technique as for the 
solar spectrum. In few cases the  S$_{\rm i}$ resulted to be significantly different from the 
O$_{\rm i}$, in particular below 5167\,\AA. This region includes the C$_2$ bands of the Swan system 
\citep{SW57} and, in particular  the one used by \citet[Table 2]{GO16} to define the C2U index, 
thus suggesting that the mismatch between S$_{\rm i}$ and O$_{\rm i}$ may be related to differences 
in the estimated and actual stellar Carbon content. The abundance determination of C is quite 
challenging and the values of [C/Fe] derived by GES are, in general, less accurate than for the 
other elements. In particular, GES does not include carbon abundances for 254 stars and, 
in almost all the other 2057 cases, the estimated [C/Fe] is based on the analysis of only 2 
(1572 stars) or even 1 (478 stars) spectral lines. To further investigate the role of [C/Fe] 
in the comparison of our synthetic spectra with the observed ones we computed for the whole sample 
of UVES-U580 stars the C2U index from both S$_{\rm i}$ and O$_{\rm i}$. Figure\,\ref{fig:C2U} shows that, 
while in most cases the observed and synthetic C2U indices agree between $\pm 0.05$\,mag, there are 
99 stars which show larger differences suggesting that their estimated (or assumed) [C/Fe] 
are not correct. As a consequence, these stars are, for precaution, removed from our UVES-U580  sample 
which, at the end, consists of 2212 stars with normalized observed spectra and NSPs. 

In order to quantitatively estimate the agreement between each pair of S$_{\rm i}$ and O$_{\rm i}$ within
the working GES sample, 
we calculate the same  figure of merit, $r^{\rm i}_{med}$, used to compare the  O$_{\rm i}$ and 
S$_{\rm i}$ spectra of the five giants in Table\,\ref{tab:giants}. The use of $r^{\rm i}_{med}$  as an 
estimate of the accuracy of our synthetic spectra requires, however, that we take into account that its value  also  
depends on the uncertainties in the GES atmospheric parameters and in the normalization procedure. 
In order to investigate these two contributions we computed for each star, in addition to  
the figure of merit derived from synthetic spectra for the {\it nominal} set of GES atmospheric 
parameters and elemental abundances (that we call {\it n}\_$r^{\rm i}_{med}$), the synthetic spectra and the figure of merit 
by adding or subtracting for each atmospheric parameter the given  1-$\sigma$  uncertainty. Note that
we have also obtained for each new S$_{\rm i}$ the corresponding normalized observed spectrum O$_{\rm i}$. 

The goodness of the GES estimates is confirmed by the general increase of $r^{\rm i}_{med}$ 
values with respect to the {\it nominal} ones when the S$_{\rm i}$s are computed considering
the parameter uncertainties. We found that the main mean increase is caused by the adoption of 
$T_{\rm eff}- \sigma_{T{\rm eff}}$  and amounts about 10\%. The effects of varying the other parameters and of 
the normalization procedure turned out to be negligible.

\begin{figure}[h!]
\includegraphics[width=\textwidth]{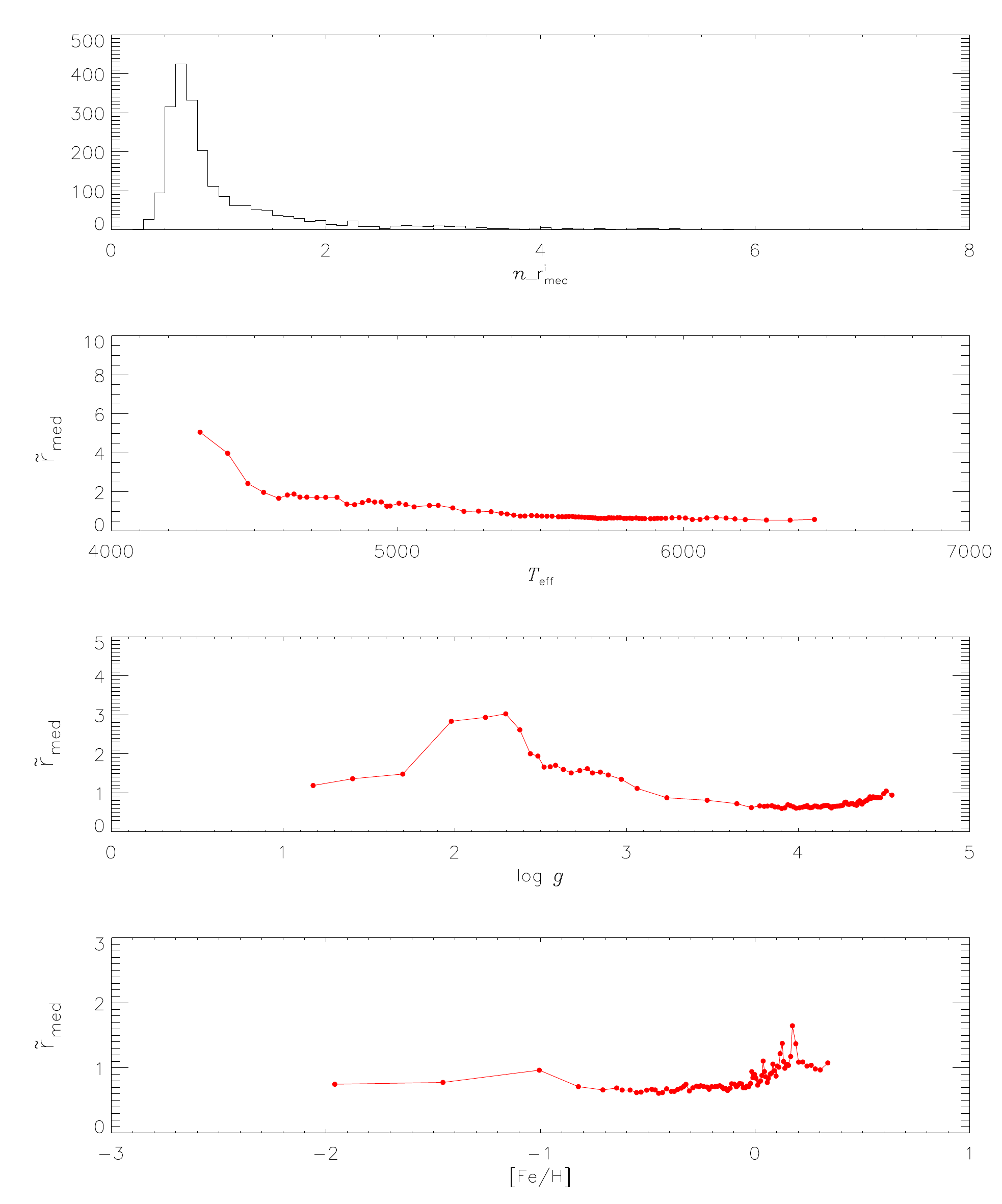}
\caption{
Distribution of {\it n}\_$r^{\rm i}_{med}$ obtained by using the {\it nominal} atmospheric parameter values in computing the synthetic spectra (top panel); 
trend  of the median values of the  {\it n}\_$r^{\rm i}_{med}$ computed in overlapped bins containing 51 stars each vs  $T_{\rm eff}$,  log\,$g$, and
[Fe/H] (second, third, and fourth panel respectively from top to bottom).
\label{fig:Rmed}}
\end{figure}

The resulting distribution of the {\it n}\_$r^{\rm i}_{med}$  is presented  in the top panel of
Figure\,\ref{fig:Rmed} which shows that the reliability of the astrophysical {\it gf}-values in our list and, 
therefore, of the resulting synthetic spectra, is validated by the small {\it n}\_$r^{\rm i}_{med}$ 
values for the bulk of the 2212 stars. In fact, as can be seen, the {\it n}\_$r_{med}$ distribution is strongly peaked at values below 1,
with the maximum of the distribution between 0.6 and 0.7, 
attesting that, in most cases, the differences between S$_{\rm i}$ and O$_{\rm i}$  are on the same 
order (or even lower) of the estimated observational errors. The presence of a wing in the distribution towards 
higher {\it n}\_$r_{med}$ values is mainly due, as expected, to the coolest stars as can be seen 
in the second panel of  Figure\,\ref{fig:Rmed} where the median values 
of the {\it n}\_$r^{\rm i}_{med}$  in partially overlapping bins containing 51 stars each 
($\tilde r_{med}$) are plotted versus $T_{\rm eff}$. Actually, for these objects, a lower accuracy of 
our synthetic spectra in reproducing the observed ones is somehow unavoidable due to both the 
difficulties in computing their atmosphere models and the absence, in our computations, of 
the tri-atomic molecular lines. In the third and fourth panels we depict the trends of 
$\tilde r_{med}$  vs log\,$g$ and $[$Fe/H$]$, respectively. The increase of $\tilde r_{med}$ for log\,$g$
between 2.0 and 2.4\,dex probably reflects the high number of cool GES stars in this gravity interval.  
The increase of  $\tilde r_{med}$ at $[$Fe/H$]\gtrsim 0$ can plausibly be attributed to the
insufficient improvement  (or lack of it) of the log({\it gf})s of weak lines in the solar and giant stellar
spectra that are more prominent in the super-metal-rich regime.\\

In conclusion we derived and validated astrophysical {\it gf}-values for 899, 77, and 1428 atomic, molecular and 
predicted lines, 
respectively. In particular, by adapting also the log({\it gf}) values of PLs, we minimized both  
the unavoidable underestimation of the blanketing in the synthetic spectra if PLs are  ignored 
(see discussion in C14)  and  the risk of  worsening the match with the observed spectrum if  
PLs with incorrect intensity are used \citep[see Figure\,3 in][]{MU05}.
Therefore, on the basis of the above  discussion, we confidently computed synthetic spectra for each 
of the ATLAS12 model listed in Section\,\ref{subsec:models} and generate the final INTRIGOSS library.

The INTRIGOSS spectral library  and  the linelist  used to compute the synthetic spectra
are available  online  together with  auxiliary data as described in Section\,\ref{sec:data}.

\section{Comparison with other spectral libraries}\label{sec:comparison}
One of the main applications of stellar spectral libraries is the automatic analysis of spectra in stellar surveys to derive  atmospheric parameters.
Several examples can be found in the literature by using different spectral libraries and numerical codes, see for example \citet{GA15}, \citet{WO16},
B17, \citet{KO17},  etc. The accuracy of the obtained atmospheric parameters depends both on the reliability of 
the input spectral libraries and on the algorithms implemented in
the numerical codes used to derive them. It is therefore necessary to remove the effect of the different parameter estimate codes if we want to compare 
the spectral libraries.
Thus we decided to use as reference the UVES-U580 stars together with the homogeneously derived set of GES atmospheric parameter values that we consider
as one of the best currently available. Therefore, the comparison of observed stellar spectra
with the theoretical predictions of any synthetic stellar libraries can be safely performed by using
these GES parameter values as input.
In order to perform such a comparison we downloaded the following  spectral libraries
available on-line: AMBRE, GES\_Grid\footnote{It should be noticed that  GES\_Grid library computed for internal GES use  is  based on the same
methodology adopted, when computing the AMBRE spectra \citep{deLA12} but with several improvements like, in particular, a more accurate linelist.}, 
PHOENIX, C14,  and B17. To make a fair comparison, since we could not compute for each UVES-U580 star the corresponding 
{\it nominal} synthetic spectra from the literature libraries, we adopted the following approach: 
within the six (j) libraries and INTRIGOSS we computed the corresponding synthetic spectrum  for each UVES-U580 star by   linearly
interpolating  in  $T_{\rm eff}$, log\,$g$, [Fe/H], $<$[$\alpha$/Fe]$>$ or [Mg/Fe], and $\xi$ at the {\it nominal} GES atmospheric parameter values.
The interpolation procedure implies the addition of (systematic) errors that will depend on  two main factors: the spacing in 
the grid nodes and the applied interpolation strategy  \citep[see for example][]{MES13}. 
Whilst a full analysis of the effects of interpolating within INTRIGOSS 
is beyond the scope of this paper, we hereafter provide some tests to estimate the errors associated with our
linear interpolation.

In order to evaluate the errors introduced by our interpolation procedure we computed by using INTRIGOSS prescriptions the intra-mesh atmosphere models and the 
corresponding synthetic spectra, NSPs and FSPs, of 50 representative UVES-U580 stars, i.e. by using their nominal GES $T_{\rm eff}$, log\,$g$, [Fe/H], $<$[$\alpha$/Fe]$>$, 
and $\xi$, but not their individual element abundances\footnote{These 50 intra-mesh synthetic spectra are also available on the 
website http://archives.ia2.inaf.it/intrigoss}. For each star we computed the mean value and the standard deviation ($\sigma_{\rm rd}$) of the  relative
differences between the interpolated and the intra-mesh spectra. The mean  relative differences can be used to evaluate the interpolation 
error introduced in the overall spectrum levels while the standard deviations can be seen as an estimate of the ``noise'' introduced point-by-point. 
In the following Sections\,\ref{subsec:NSPs} and  \ref{subsec:FSPs} we will use these values to provide estimates of the interpolation errors introduced 
by interpolating NSPs and FSPs, respectively.

\subsection{The normalized synthetic spectra, NSPs} 
\label{subsec:NSPs}
We used the synthetic spectra obtained by interpolating the different j spectral libraries to normalize the corresponding UVES-U580 spectra 
and to compute the   $r^{j}_{med}$  figure of merit as described in Section\,\ref{subsec:spectra}.
Due to the different wavelength and parameter space coverage of the six libraries, namely INTRIGOSS, AMBRE, GES\_Grid, PHOENIX, C14,  and B17,
we restricted our analysis to the regions in common,
i.e. 4900-5370\,\AA, and -1.0$\leq$\,[Fe/H]\,$\leq$\,0.5\,dex, but for C14 which is limited to [Fe/H]\,$\leq$\,0.2\,dex.

\begin{deluxetable*}{cccccccc}[ht!]
\tablecaption{Median values of the normalized $r_{med}$s for different spectral libraries \label{tab:medR}}
\tablecolumns{8}
\tablewidth{0pt}
\tablehead{
\colhead{Library} & \colhead{INTRIGOSS$_{{\rm Mg}}$} & \colhead{INTRIGOSS$_{\alpha}$} & \colhead{AMBRE} & \colhead{GES\_Grid}& \colhead{PHOENIX} & \colhead{C14} & 
\colhead{B17}  \\
}
\startdata
 $R_{med}$ &    1.043 &  1.025 &   1.313 &   1.266 &   2.161 &   1.700 &  1.298  \\
 $\sigma$ &    0.003 &  0.003 &   0.021 &   0.011 &   0.033 &   0.021 &  0.014 \\
\enddata
\end{deluxetable*}

\begin{figure*}[h!]
\includegraphics[width=\textwidth]{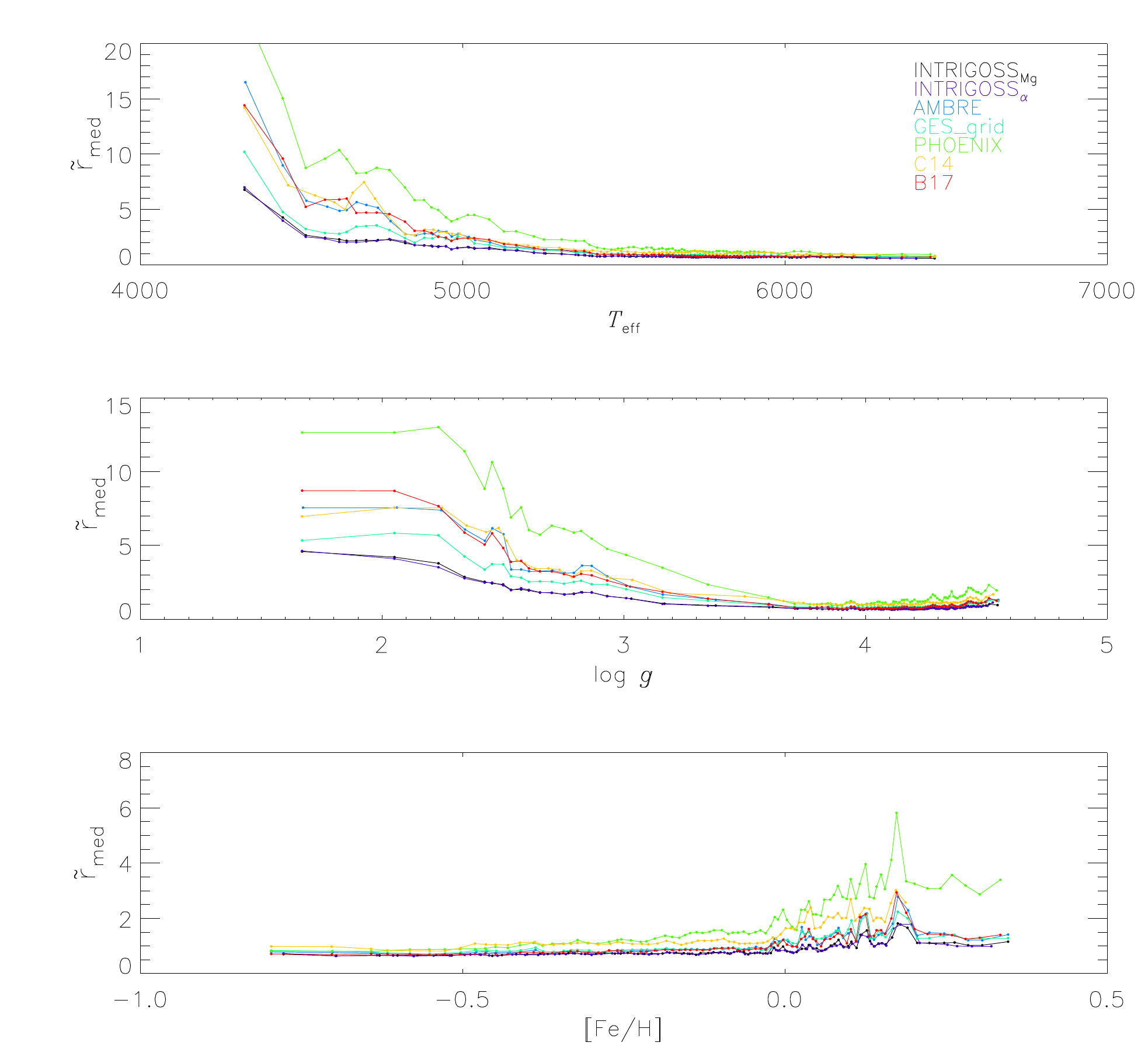}
\caption{Trends of the median values of the computed $r^{\rm j}_{med}$s for the different spectral libraries (identified by different colours)
versus GES  $T_{\rm eff}$, log\,$g$, and [Fe/H] (top to bottom); see text and Figure\,\ref{fig:Rmed}.
\label{fig:Rint}}
\end{figure*}

\begin{figure}[h!]
\includegraphics[width=13cm]{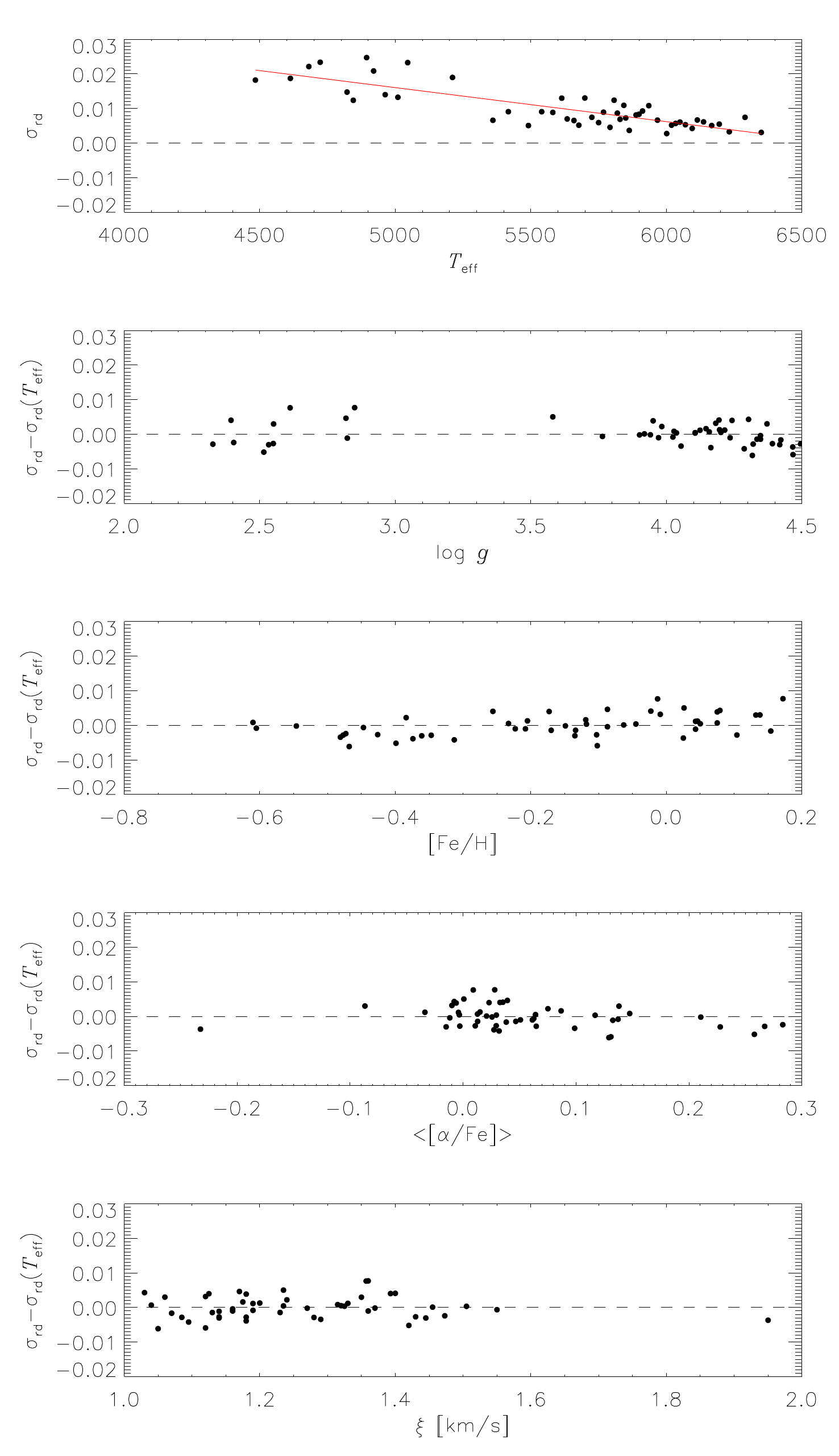}
\caption{Trends of the standard deviation ($\sigma_{\rm rd}$) of the relative
differences between the 50 interpolated and intra-mesh NSPs versus  $T_{\rm eff}$, log\,$g$, [Fe/H], $<$[$\alpha$/Fe]$>$, 
and $\xi$. The general trend of $\sigma_{\rm rd}$  versus  $T_{\rm eff}$, $\sigma_{\rm rd}(T_{\rm eff})$, as derived by a linear regression 
(red line in the top panel), has been subtracted from $\sigma_{\rm rd}$ in the other panels.
\label{fig:interp}}
\end{figure}

Figure\,\ref{fig:Rint} shows the trend of the  $\tilde r^{j}_{med}$ for the different spectral libraries
versus GES  $T_{\rm eff}$, log\,$g$, and [Fe/H]. 
Table\,\ref{tab:medR} summarizes, for each $j$-spectral library,
the variations of  $r^{j}_{med}$s with respect to the {\it nominal} ones, as evaluated by computing $R^{j}_{med}$=median($\frac{r^{j}_{med}}{n\_r_{med}}$), i.e. 
the median value of the normalized $r_{med}$s.
As can be seen, the  use of synthetic spectra computed at the 
$T_{\rm eff}$, log\,$g$, [Fe/H], [Mg/Fe] or $<$[$\alpha$/Fe]$>$, and $\xi$ by interpolating the INTRIGOSS library leads to a general increase of 
 a few percents of their $r_{med}$s
with respect of the corresponding {\it n}\_$r_{med}$s thus confirming that
the best agreement with the observed spectra can be reached, in general,  by using ad-hoc models and  the individual element abundances instead of
the average metallicity and the average abundance ratio of the $\alpha$-elements. 
Table\,\ref{tab:medR} shows that slightly better results are obtained by interpolating according to the $<$[$\alpha$/Fe]$>$ stellar value instead
 of using [Mg/Fe] indicating that, even if one
of the major contributors to absorption in this wavelength range is  Mg (and MgH), also the behaviour of all the other $\alpha$-elements plays a not negligible
role and must be properly taken  into consideration. 
 The increase of  $r_{med}$s also contains the 
contribution introduced by the interpolation among the INTRIGOSS grid nodes.
By comparing the  50 intra-mesh normalized synthetic spectra (see Section\,\ref{sec:comparison}) with the corresponding interpolated ones,
we found that the mean  relative differences are on the order of $\pm 0.1$\% showing that the interpolation does not introduce
significant inaccuracies in the overall normalized spectrum levels. 
As far as the relative standard deviations are concerned, top panel of Figure\,\ref{fig:interp} shows that $\sigma_{\rm rd}$ increases 
with decreasing temperature showing that for  $T_{\rm eff}\gtrsim 5500$
interpolation errors become visible when working with spectra
at SNR\,$\gtrsim$\,100 ($\sigma_{\rm rd}\simeq0.01$)  while for lower temperatures  interpolation errors become significant also for SNR\,$\sim$\,50 ($\sigma_{\rm rd}\simeq0.02$).
Furthermore, from Figure\,\ref{fig:interp} we can also see that for the 
other parameters, once the general trend of the standard deviations with  $T_{\rm eff}$ is removed 
(linear fit shown in the first panel of Figure\,\ref{fig:interp}), the expected interpolation errors are less 
than 1\% and do not show any significant trend with regard to the parameter values. 
On the basis of these results we plan to use a finer step in  $T_{\rm eff}$ in the future extended version of INTRIGOSS in order to reduce interpolation errors.

As far as the other spectral library are concerned,  we point out  that the $\tilde r^{j}_{med}$  in  Figure\,\ref{fig:Rint} and the 
 $R_{med}$s in  Table\,\ref{tab:medR}  
obtained with AMBRE, GES\_Grid, PHOENIX, C14,  and B17 libraries are much larger than those derived by using INTRIGOSS interpolated spectra
(both INTRIGOSS$_{Mg}$ and INTRIGOSS$_{\alpha}$). Their values include both the interpolation errors and the effects of the differences in
physical assumptions and  atomic and molecular line data in the different libraries. 
Unfortunately, the quantitative estimates of the uncertainties introduced by the interpolation cannot easily be given because we do not have at our disposal
the equivalent intra-mesh synthetic spectra for  AMBRE, GES\_Grid, PHOENIX, C14,  and B17 libraries. Furthermore, different libraries have 
different nodes, sometimes not equally spaced, and these differences may reflect in different interpolation errors.
Nevertheless,  it is very unlikely that such large $R_{med}$ values as those reported in Table\,\ref{tab:medR} can be due only
to the interpolation errors. Thus, we conclude that the differences in the $R_{med}$ values indicates that INTRIGOSS spectra
better reproduce the observed spectra of our UVES-U580 sample than the synthetic spectra from the other libraries.

The better performance of INTRIGOSS synthetic spectra can  be inferred   not only by the low $R_{med}$ values in Table\,\ref{tab:medR}  but also
by the much smaller spreads of their normalized $r_{med}$ values  which reach a maximum of $\frac{r^{j}_{med}}{n\_r_{med}}$=1.6  with only the 10\% of points above 1.2. 
On the other hand, not only the $R_{med}$ values for the other five libraries are higher but also the spreads of the
normalized  $r^{j}_{med}$s span interval several units wide with 10\% of the $\frac{r^{j}_{med}}{n\_r_{med}}$ values above 2.2, 1.9, 4.3, 2.6, 2.2 for 
 AMBRE, GES\_Grid, PHOENIX, C14,  and B17, respectively.  In particular, inspection of
Figure\,\ref{fig:Rint} shows that the coolest and/or metal richest stars are those  characterized  by higher $\tilde r^{j}_{med}$ values
thus confirming that they are the most critical objects.

\subsection{The surface flux spectra, FSPs}
\label{subsec:FSPs}

\subsubsection{Comparison of FSPs versus observed SEDs} \label{subsubsec:seds}
\begin{figure*}[h!]
\includegraphics[width=\textwidth]{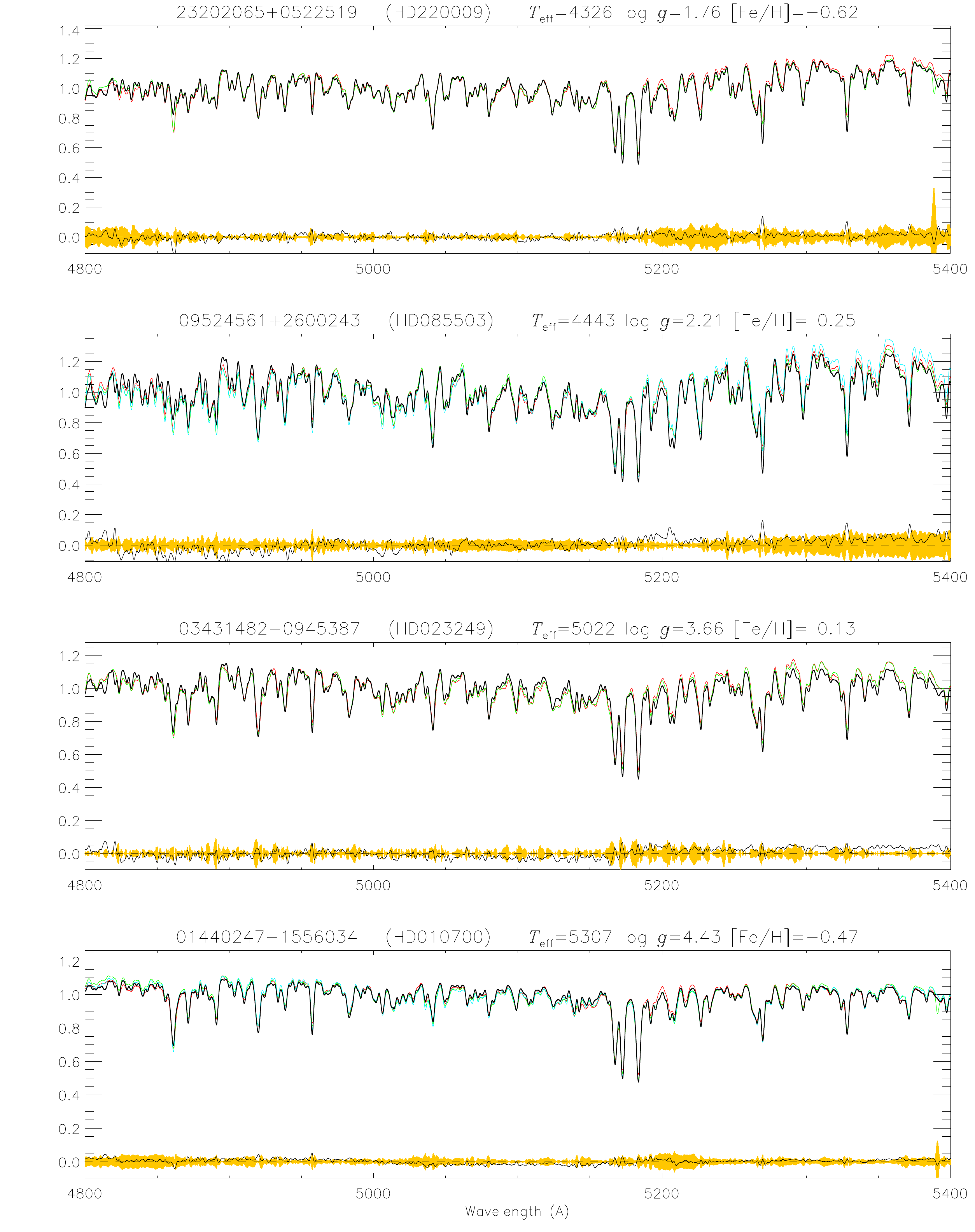}
\caption{Comparison of flux calibrated spectra (MILES in red, Indo-US in green, and ELODIE in light blue)
scaled according to their median values and the corresponding {\it n}\_FSPs (black). 
Black curve at the bottom of each panel shows the flux difference, i.e.  the average of the observations  minus the
correspondent {\it n}\_FSP, superimposed on the 3$\sigma$ of the observed spectra (yellow area). 
\label{fig:SEDs1}}
\end{figure*}

\begin{figure*}[h!]
\includegraphics[width=\textwidth]{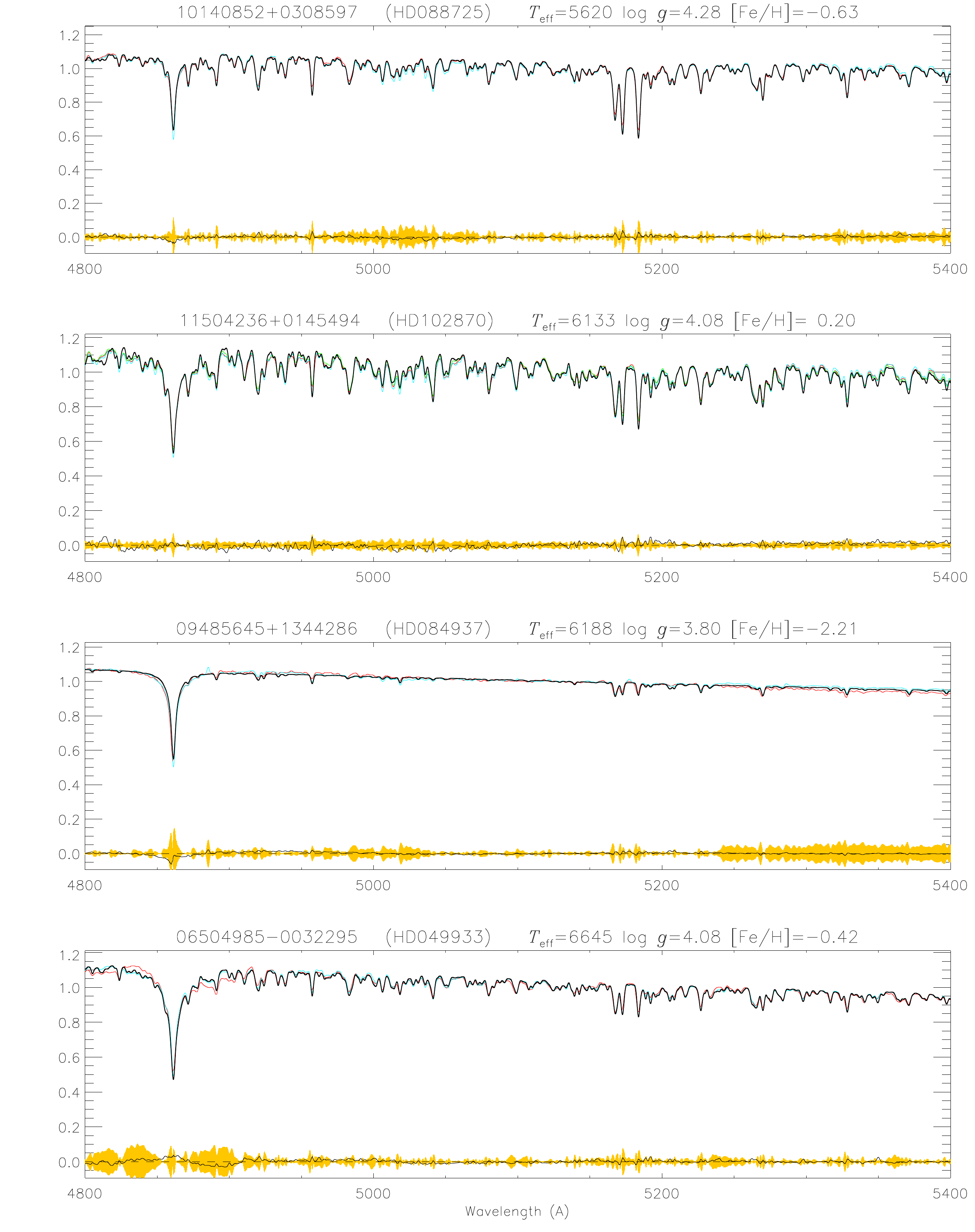}
\caption{As in Figure\,\ref{fig:SEDs1}.
\label{fig:SEDs2}}
\end{figure*}

In this Section, the INTRIGOSS FSPs are compared to observed flux calibrated spectra.
A search for stars of our UVES-U580 sample within  the  ELODIE  \citep{PRU01,PRU07}, INDO--U.S. \citep{VA04},  and MILES \citep{SA06} SED libraries
 provided a list of about 20 stars in common.
Eight of them are present in MILES and, at least, in one other SED library  and can be used to check the predictions of INTRIGOSS as far as the FSPs are concerned.
We chose MILES as the reference SED source because it is one of the most used standard empirical library for 
stellar population  models  \citep[see for example][]{VA15}.
Since  only relative fluxes are, in general, needed for this kind of studies we did not attempt to use absolute fluxes but
we scaled the observed stellar SEDs and the corresponding synthetic 
spectra computed using the {\it nominal} GES atmospheric parameter values and the individual
element abundances of each star (hereafter  {\it n}\_FSPs) according to their median flux value.
In Figures\,\ref{fig:SEDs1} and \ref{fig:SEDs2} we plot the  scaled {\it n}\_FSPs and observed SEDs together with the 
residuals obtained after computing the average of the available SEDs, i.e.  $<$SED$>$-{\it n}\_FSP,  and the $3\sigma <$SED$>$ uncertainties. These two figures 
indicate that the {\it n}\_FSPs for the stars at $T_{\rm eff}<5300$\,K reproduce, without any systematic trend,  the mean observed SEDs within $\sim3$\%, 
while for higher $T_{\rm eff}$s the agreement is within 1\%.
We also remark, by looking at Figures\,\ref{fig:SEDs1} and \ref{fig:SEDs2},  the absence of the excessive opacity near 
5200\,\AA{} found  by  C14 in her synthetic spectra (see their Figure\,10). We can conclude that the {\it n}\_FSPs of the stars in  Figures\,\ref{fig:SEDs1} and \ref{fig:SEDs2}
accurately predict the observed SEDs. Unfortunately, the small number of UVES-U580 stars with accurate observed SEDs 
does not allow us to check in details, through a complete coverage and a fine sampling, the accuracy of FPSs over the whole extension of INTRIGOSS library 
in the atmospheric parameter space. Therefore, in next Section we will use, instead of SEDs, a different approach based on the comparison of spectral 
feature indices.
   
\subsubsection{Comparison of observed and synthetic Mg$_1$, Mg$_2$, and Mgb Lick/SDSS indices}
For a long time (and still now), several spectroscopic analyses of stellar populations have relied on the Lick/IDS system of indices \citep{GO93, 
WR94, WR97, TH04, KO05, WR14}. More recently, several authors introduced new Lick-like systems  \citep[see for example][]{KI16} to avoid the possible
uncertainties associated with the response curve of the original Lick/IDS spectrograph and/or  any potential loss of information 
that would occur in degrading spectra obtained from  current surveys at medium resolution (e.g. R$\sim$1800) like the Sloan Digital Sky Survey, 
SDSS, \citep{YOR00} or the Large Sky Area Multi--Object Fiber Spectroscopic Telescope survey, LAMOST,\footnote{http://www.lamost.org/LAMOST})
to match the low resolution (R$\sim$630) of the original Lick/IDS system. One of these Lick-like systems is 
the Lick/SDSS  \citep{FR10} which was built from observed energy distributions, SEDs, at R=1800 and which is not affected by any particular instrumental signature.
The  Lick/SDSS indices are computed by integrating the spectrum in  central bandpasses covering prominent stellar features after normalization to a pseudo-continuum
defined via two bracketing blue and red side bands, and are, therefore, not very sensitive to small inaccuracies in the flux spectra calibration. Nevertheless,
since the three bandpasses cover, in some cases, relatively large wavelength ranges, some indices are sensitive not only to the main absorption feature they were designed to measure,
but also to the overall line blanketing present in the spectra.
It is therefore possible to use them to check the accuracy and completeness of the  atomic, molecular, and predicted lines used to compute the FSPs.
In fact, if it turns out that the FSPs are able  to accurately predict the observed  indices, then the accuracy  of the atomic and molecular absorption caused
by the atmospheric models used to derive them, and therefore of the FSPs themselves, would be substantiated. In the following we will use the comparison
between observational and synthetic Lick/SDSS indices as a method to evaluate the validity of the FSPs over the full atmospheric parameter space covered by the UVES-U580 stars.
In particular, we chose the Mg$_1$, Mg$_2$, and Mgb indices which are characterized by quite extended bandpasses  falling in the wavelength region covered by our synthetic spectra.
First we computed the observational indices, hereafter UVES-U580\,indices, for each of the 2212 stars of the UVES-U580 sample,
after removing from the observed (stacked) UVES-U580 spectrum, degraded at  $R=$1800, 
the instrumental signature by means of the corresponding {\it n}\_FSP.
Then, we computed the corresponding synthetic indices from INTRIGOSS FSPs and from the spectra libraries listed in Section\,\ref{subsec:NSPs}
with the same interpolation adopted in Section\,\ref{subsec:NSPs}. 
Unfortunately, the GES\_Grid, and B17 libraries contains only normalized spectra and cannot be used to 
compute Lick/SDSS indices.  Therefore, we were able to compute the following synthetic indices:
\begin{itemize}
\item {\it n}\_FSP indices  obtained from the above-defined {\it n}\_FSPs;
\item {\it Interp}\_FSP$_{\alpha}$ indices  obtained from the spectra computed by interpolating INTRIGOSS FSPs at the stellar $T_{\rm eff}$, log\,$g$, [Fe/H],  $\xi$, 
and $<$[$\alpha$/Fe]$>$;
\item {\it Interp}\_FSP$_{\rm Mg}$ indices obtained from the spectra computed by interpolating INTRIGOSS FSPs at the stellar [Mg/Fe] abundance ratio instead of at $<$[$\alpha$/Fe]$>$;
\item {\it Interp}\_PHOENIX indices obtained from the spectra computed by interpolating the flux calibrated PHOENIX spectra at the stellar $T_{\rm eff}$, log\,$g$, [Fe/H], and 
$<$[$\alpha$/Fe]$>$;
\item {\it Interp}\_C14 indices obtained from the spectra computed by interpolating the flux calibrated C14 spectra at the stellar $T_{\rm eff}$, log\,$g$, [Fe/H], and 
$<$[$\alpha$/Fe]$>$.
\item {\it Interp}\_AMBRE indices obtained from the spectra computed by interpolating the flux calibrated AMBRE spectra at the stellar $T_{\rm eff}$, log\,$g$, [Fe/H], and 
$<$[$\alpha$/Fe]$>$.
\end{itemize}

\begin{figure*}[ht!]
\includegraphics[width=\textwidth]{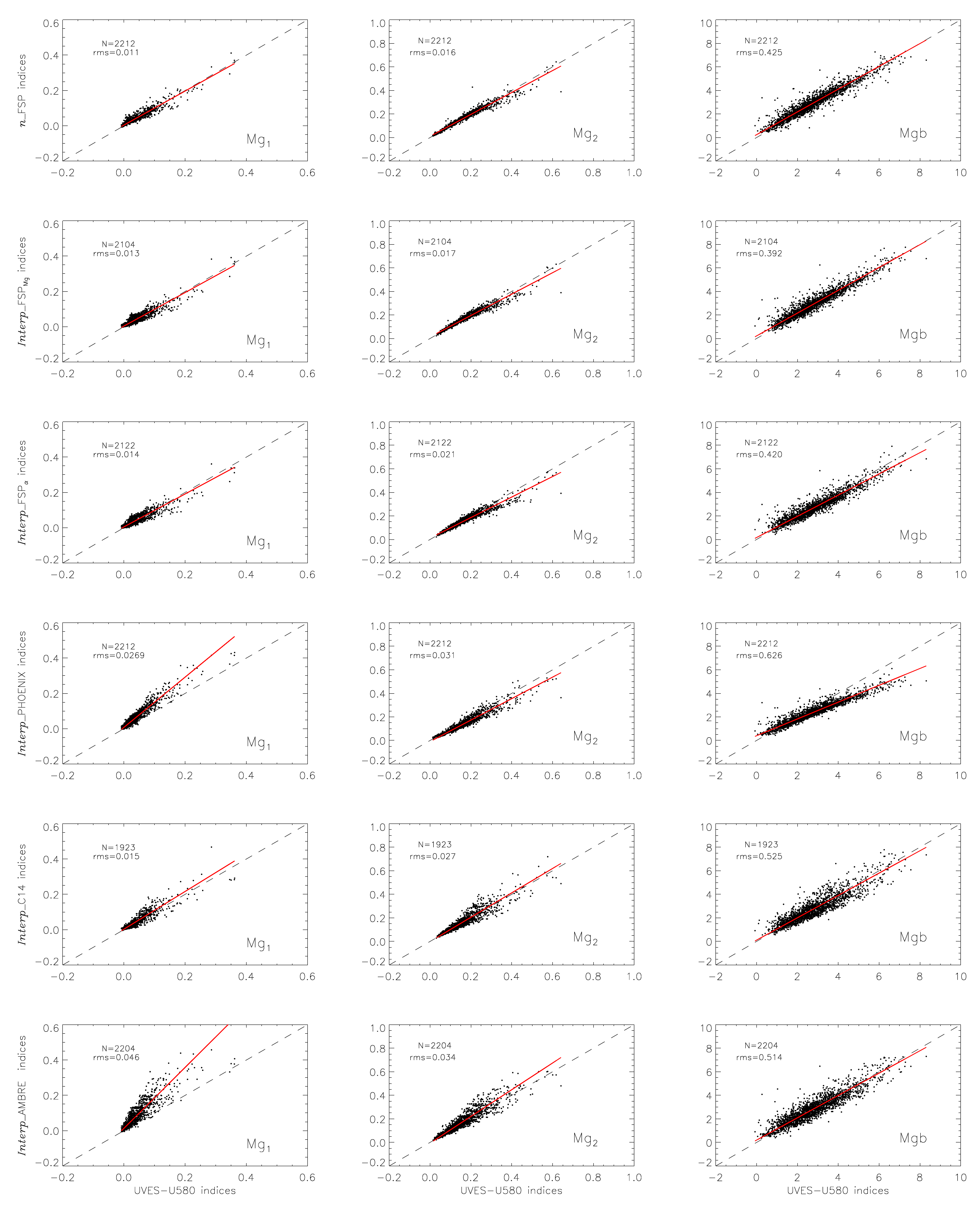}
\caption{Comparison of synthetic  and observational Lick/SDSS indices Mg$_1$  (left panels), Mg$_2$  (central panels), and Mgb (right panels).
Indices computed from synthetic spectra are plotted versus those measured from UVES-U580 spectra (see text) together with the 45$^{\circ}$ (dashed) and the regression (red) lines. 
From top to bottom:  {\it n}\_FSP indices  computed from {\it n}\_FSPs;  {\it Interp}\_FSP$_{\rm Mg}$ and {\it Interp}\_FSP$_{\alpha}$  indices computed from INTRIGOSS; {\it Interp}\_PHOENIX,
 {\it Interp}\_C14, and {\it Interp}\_AMBRE indices computed from PHOENIX, C14, and AMBRE libraries respectively (see text).
\label{fig:indici}}
\end{figure*}

While we were able to compute {\it n}\_FSPs and, therefore the corresponding  {\it n}\_FSP indices, for all the stars in the UVES-U580 sample,
the number of stars for which the interpolation within the spectral grids was possible varies because of the different  parameter space coverages. In Figure\,\ref{fig:indici}
we plot the synthetic indices versus the UVES-U580\,indices. Each panel contains  the number of stars (N), the rms of the 
deviations from the 45$^{\circ}$ line and
the synthetic vs UVES-U580 regression lines. It can be seen that there is a very good agreement between {\it n}\_FSP indices and UVES-U580 ones indicating that the 
blanketing of the FSP spectra correctly predict the observed one. The non significant increase of the rms   when using {\it Interp}\_FSP$_{\alpha}$ and {\it Interp}\_FSP$_{\rm Mg}$ indices
indicates that spectra computed from individual star models and element abundances provides synthetic indices that are almost equivalent to those obtained by 
{\it Interp}\_FSPs. This different result with respect to that one obtained by looking at the $r_{med}$ values (see Section\,\ref{subsec:NSPs}) could be ascribed
to the lower sensitivity of the spectral index comparison with respect to that one performed  at each wavelength point. 

The panels referring to {\it Interp}\_PHOENIX, {\it Interp}\_C14, and {\it Interp}\_AMBRE indices
show a larger rms and/or some systematic trend. The deviations of the Mg$_1$, Mg$_2$ indices derived from PHOENIX and AMBRE spectra from the 45$^{\circ}$ line 
indicates significant differences in the blanketing predicted by these spectra. In the case of the indices derived from 
C14 spectra the deviations are smaller and the increase of rms is present only in Mg$_2$ and Mgb. It is worthwhile to recall that the
comparison with the UVES-U580 indices is, in this case, limited to stars with $-1.0<[Fe/H]<0.2$\,dex due to the absence of super-metal rich spectra in the C14 library.
The small systematic deviations  may indicate that C14 has, on average, more accurate line lists than PHOENIX and that the increase in rms with respect
to the upper panels may be due to the different treatment of the PLs.

\begin{figure*}[ht!]
\includegraphics[width=\textwidth]{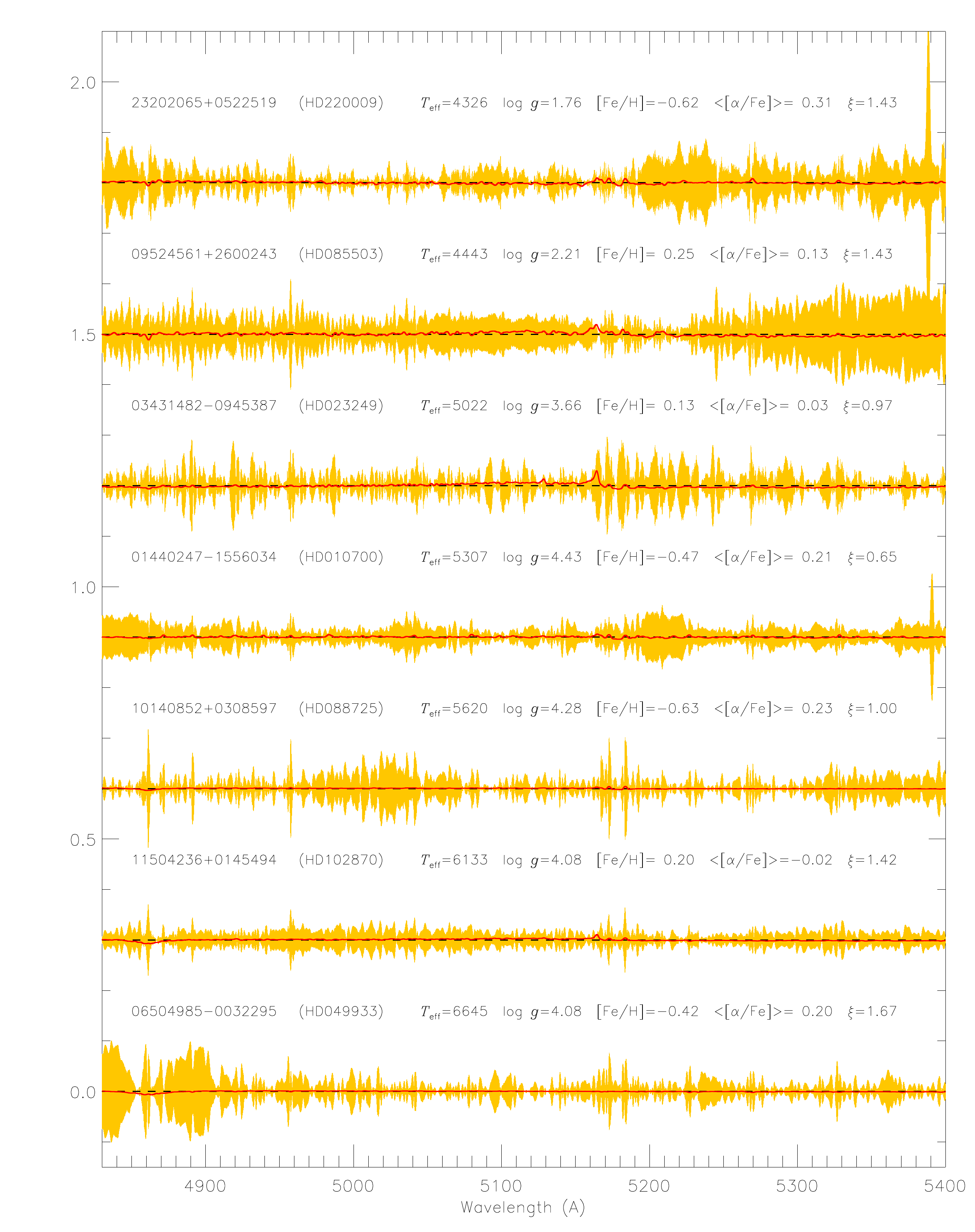}
\caption{Comparison  between the  3$\sigma$ of the $<$SEDs$>$ (yellow area) and the differences between the intra-mesh and the interpolated
INTRIGOSS FSPs (red lines) for seven of the stars discussed in Section\,\ref{subsec:FSPs}. Horizontal
dashed black lines represent the zero levels after vertical shifts applied to better visualize all the stars in the same plot.
\label{fig:miles_interp}}
\end{figure*}

Concerning the errors introduced by our interpolation procedure, we confirm the results already obtained in Section\,\ref{subsec:NSPs} also for FSPs.
In addition,  we computed intra-mesh FSPs for the seven stars in Figures\,\ref{fig:SEDs1} and \ref{fig:SEDs2} for which we were able to compute
interpolated INTRIGOSS FSPs\footnote{We cannot compute the interpolated synthetic spectrum for the  09485645+1344286 UVES-U580 star 
 (HD084937) since its [Fe/H] is -2.21\,dex.}. 
Then,
we compare the $3\sigma$   uncertainties of the $<$SED$>$ of these stars 
with the differences between the intra-mesh  and the interpolated INTRIGOSS FSPs.  
Figure\,\ref{fig:miles_interp} shows that the interpolation procedure introduces inaccuracies in the spectra (red lines)
which are well below the SEDs  uncertainties (yellow areas). 

Eventually, we compared  the  {\it Interp}\_FSP$_{\alpha}$ indices  with those computed by using the 50 intra-mesh FSPs 
to estimate the effect of the interpolation on the computation of the indices 
and we did not find any systematic trend in the differences. The use of the interpolated spectra instead of
the intra-mesh ones introduces an rms scatter which is one order of magnitude smaller than those reported in 
the first row of Figure\,\ref{fig:indici} showing that the interpolation error on the indices does not undermine
the above-given discussion about the indices computed with the different spectral libraries.

In conclusion, the comparison of observational and synthetic Mg$_1$, Mg$_2$, and Mgb Lick/SDSS indices indicates that the INTRIGOSS FSPs predict well the observed blanketing
thus suggesting that this library can provide accurate synthetic SEDs not only for the stars discussed in Section\,\ref{subsubsec:seds} but also for all of those 
in the UVES-U580 sample.

\section{Data products} \label{sec:data}
The INTRIGOSS spectral library   and  the linelist  used to compute the synthetic spectra are  available  on the website 
http://archives.ia2.inaf.it/intrigoss together with  auxiliary data. 

The  synthetic spectra are computed from  4830 to 5400\,\AA~at wavelength sampling 
$\Delta \lambda = 0.01$\,\AA, rotational velocity of 0\,km\,s$^{-1}$ , and, 
in order to be consistent with the ATLAS12 models, with microturbulent velocities $\xi$ = 1 and 2\,km\,s$^{-1}$  
leading to a final total number of 7616 NSPs and 7616 FSPs. 

The gaps in the final grid are due to the absence of converging ATLAS12 atmosphere models for  log\,$g=0.5$\,dex 
and $T_{\rm eff}\geq6250$\,K. In order to keep the grid homogeneous, we decided to avoid any patches 
based on, for example, the use of ATLAS9 atmosphere models. Work is in progress to attain convergence
of  ATLAS12 code at relatively high temperatures and low surface gravities. 

The INTRIGOSS spectra are provided in FITS binary table format and can be downloaded by selecting:
\begin{itemize}
 \item the type of spectra: NSP, FSP, or both;
 \item a range in $T_{\rm eff}$ and/or log\,$g$ and/or  [Fe/H] and/or [$\alpha$/Fe] and/or $\xi$, or the whole library (15232 spectra);
\end{itemize}

The following auxiliary data are also available from links given  in the http://archives.ia2.inaf.it/intrigoss-details web-page:
\begin{itemize}
 \item A FITS binary table with the line list of atomic and molecular transitions used in computing INTRIGOSS synthetic spectra.
       The table contains 1427628 entries in the format of the linelist file used by SPECTRUM code \citep[see Section 3.3.1 and 3.6 of Documentation for SPECTRUM][]{GR94},
       i.e, for each line we list:
\begin{trivlist}
        \item WAVELENGTH: wavelength in \AA;
	\item ELEM\_ION: element and ion identifier, e.g. 26.1 for FeII;
        \item ISOTOPE: mass number of isotope (the 0 code corresponds to entries representing all possible isotopes for that species taken together); 
        \item ELOW: energy of the lower state in cm$^{-1}$; 
        \item EHIGH: energy of the upper state in cm$^{-1}$; this entry is sometimes used to encode the molecular band information since only ELOW is used in molecular
              calculation;
        \item LOG\_GF: the logarithm of the product of the statistical weight of the lower level and the oscillator strength for the transition; 
        \item FUDGE: a fudge factor to adjust the line broadening;
        \item TRANSITION\_TYPE: the type of transition;
        \item REFERENCE: The sub-set of lines with derived astrophysical {\it gf} values are indicated  as FRA18 and FRA18\_P for laboratory and 
       predicted lines, respectively.
\end{trivlist}    
 \item A FITS binary image with the very high SNR ($\sim$4000) observed solar spectrum described in Section\,\ref{subsubsec:sun} and 
    used to derive astrophysical log({\it gf}) values;
\item A set of 100  FITS binary tables with synthetic spectra (50 NSPs and 50 FSPs) computed  at 50 representative intra-mesh positions. 
      These spectra are used in Section\,\ref{subsec:NSPs} to test the errors arising from our interpolation procedure of the INTRIGOSS spectra.  
\end{itemize}

\section{Summary} \label{sec:summary}

In this paper we present a new high resolution synthetic spectral library, INTRIGOSS, which covers the parameter space range of F, G, and K stars. INTRIGOSS 
is based  on  atmosphere models computed with ATLAS12 which allowed us to specify each individual element abundance.
The normalized (NSP) and surface flux (FSP) spectra, in the 4830-5400\,\AA~ wavelength range, were computed in a fully consistent mode by means of 
SPECTRUM v2.76f code using  the detailed solar composition 
by \citet{GRE07} and varying it by adopting different [$\alpha$/Fe] abundance ratios. Particular attention was devoted to derive  astrophysical {\it gf}-values
by comparing synthetic prediction with a very high SNR solar spectrum and good SNR UVES-U580 spectra of cool giants. The validity of the obtained spectra and, in particular,
of the used astrophysical {\it gf}-values, was assessed by using as reference more than 2000 stars with homogeneously and accurately derived atmospheric parameter values and
detailed chemical compositions.

The greater accuracy of INTRIGOSS NSPs  with respect to other publicly available stellar libraries, i.e. AMBRE, GES\_Grid, PHOENIX, C14,  and B17, in reproducing the observed
spectra was shown by computing a figure of merit, $r_{med}$, to evaluate the consistency  of the prediction of different libraries with respect to the
spectra of the UVES-U580 sample stars.

As far as the FSPs are concerned, the comparison with  SEDs derived from   ELODIE, INDO--U.S.,  and MILES libraries showed that they reproduce the observed flux distributions
within a few percent without any systematic trend. A check on the predicted blanketing of FSPs and, therefore, on the adopted line lists (including the treatment of the Predicted
Lines) based on Lick/SDSS indices comparison confirmed the reliability of INTRIGOSS surface flux spectra and their ability to better reproduce the observational index values
with respect to PHOENIX, AMBRE and C14 libraries.

The results of the validity checks on both INTRIGOSS normalized (NSP) and surface flux spectra (FSP) make us confident that the presented spectral library, which  is
available on the web together with the adopted line list and the observed solar spectrum  (see Section\,\ref{sec:data}), will provide to the astronomical 
community a valuable tool 
for both stellar atmosphere parameter determinations and stellar population studies.

\acknowledgments

This work is based on data products from observations made with ESO Telescopes at the La Silla Paranal Observatory under programme ID 188.B-3002. 
These data products have been processed by the Cambridge Astronomy Survey Unit (CASU) at the Institute of Astronomy, 
University of Cambridge, and by the FLAMES/UVES reduction team at INAF/Osservatorio Astrofisico di Arcetri. 
These data have been obtained from the Gaia-ESO Survey Data Archive, prepared and hosted by the Wide Field Astronomy Unit, Institute for Astronomy, 
University of Edinburgh, which is funded by the UK Science and Technology Facilities Council.
This work was partly supported by the European Union FP7 programme through ERC grant number 320360 and 
by the Leverhulme Trust through grant RPG-2012-541. We acknowledge the support from INAF and Ministero dell' Istruzione, 
dell' Universit\`a e della Ricerca (MIUR) in the form of the grant ``Premiale VLT 2012''. The results presented here benefit from 
discussions held during the Gaia-ESO workshops and conferences supported by the ESF (European Science Foundation) through the GREAT Research Network Programme.
This work received partial financial support
from PRIN MIUR 2010--2011 project ``The Chemical and dynamical Evolution of the Milky Way
and Local Group Galaxies'', prot. 2010LY5N2T and  by the National Institute for Astrophysics (INAF) through the grant PRIN-2014 (``The Gaia-ESO Survey''). 
MC thanks financial support from CONACyT grant CB-2015-256961.
This research uses the facilities of the Italian Center for Astronomical Archive (IA2) operated by INAF.

\vspace{5mm}
\facilities{VLT:Kueyen}
\software{SPECTRUM  \citep[v2.76f;][]{GR94}, ATLAS12 \citep{KU05a}, LineSearcher \citep{SO15} }

\clearpage

\appendix
\section{Unidentified spectral features} \label{app:unident}

Together with the accurate determination of astrophysical log({\it gf}) values, the analysis of the Solar spectrum and of the GES spectra of the five giant stars
in Table\,\ref{tab:giants} allows us to determine where the adopted line list is missing features. 
To construct the list of these unidentified features we computed  the difference between 
the observed and synthetic spectrum of the Sun and of the five giants and the corresponding standard deviations ($\sigma{_{\rm Sun}}$, $\sigma_{\rm i}$). 
Then, we run the {\it LineSearcher} code, which is a derivation from the ARES code \citep{SO15}, using as input the six files of the differences and we obtained 
for  all the {\it absorption features} in the differences, i.e. those below zero, their center wavelength and depth.
Eventually,  we extracted from the solar  {\it LineSearcher} output  all the features which have a depth larger than
2$\sigma{_{\rm Sun}}$ while a 3$\sigma_{\rm i}$ threshold was used in the case of the five giants, due to the lower resolution and SNR of their spectra. 
At the end we checked if the so selected features were present in more than one spectra but with smaller depths.
Table\,\ref{tab:lines}  lists the wavelength of each unidentified feature together with its depth in  all the stars in which it is detectable while Table\,\ref{tab:nfeature}
shows the total number of detected unidentified features in each star spectrum (column\,2), and the number of those with depth between 2$\sigma$ and 3$\sigma$ (column\,3) and 
larger than  3$\sigma$ (column\,4). 

\clearpage

\startlongtable
\begin{deluxetable*}{ccccccc}
\tabletypesize{\small}
\tablecaption{Unidentified spectral lines\label{tab:lines}}
\tablecolumns{7}
\tablewidth{0pt}
\tablehead{
\colhead{} & \colhead{SUN} & \colhead{00241708-7206106}  & \colhead{00251219-7208053} & \colhead{00240054-7208550}   & \colhead{02561410-0029286} & \colhead{03173493-0022132} \\[-5pt]
\colhead{$\sigma$:} & \colhead{0.03} & \colhead{0.04}  & \colhead{0.04} & \colhead{0.04}   & \colhead{0.03} & \colhead{0.03} \\[5pt]
\tableline
\colhead{Wavelength [$\AA$]} & \colhead{Depth} & \colhead{Depth}  & \colhead{Depth} & \colhead{Depth}   & \colhead{Depth} & \colhead{Depth} {}
}
\startdata
            4834.60 &               0.08 &               0.12 &               0.04 &               0.10 &               0.05 &{\bf               -- } \\[-4pt]
            4844.19 &{\bf               -- } &{\bf               -- } &{\bf               -- } &{\bf                0.14 } &{\bf               -- } &{\bf               -- } \\[-4pt]
            4855.20 &               0.05 &{\bf               -- } &{\bf                0.17 } &{\bf               -- } &{\bf               -- } &{\bf               -- } \\[-4pt]
            4858.14 &{\bf                0.11 } &{\bf               -- } &               0.08 &{\bf               -- } &               0.06 &{\bf               -- } \\[-4pt]
            4861.95 &               0.08 &{\bf                0.20 } &{\bf                0.21 } &{\bf                0.18 } &{\bf                0.15 } &{\bf                0.11 } \\[-4pt]
            4866.72 &               0.04 &{\bf               -- } &{\bf               -- } &{\bf               -- } &{\bf               -- } &{\bf                0.09 } \\[-4pt]
            4880.97 &               0.06 &               0.08 &{\bf               -- } &               0.09 &               0.07 &               0.04 \\[-4pt]
            4884.94 &{\bf                0.13 } &{\bf                0.16 } &{\bf                0.21 } &{\bf                0.15 } &{\bf                0.11 } &{\bf                0.10 } \\[-4pt]
            4906.80 &               0.07 &{\bf               -- } &               0.08 &               0.05 &{\bf               -- } &               0.03 \\[-4pt]
            4916.23 &               0.07 &               0.03 &               0.06 &               0.03 &{\bf               -- } &{\bf               -- } \\[-4pt]
            4916.49 &{\bf                0.14 } &               0.04 &               0.10 &               0.06 &{\bf               -- } &{\bf               -- } \\[-4pt]
            4921.85 &{\bf               -- } &{\bf               -- } &{\bf               -- } &{\bf               -- } &{\bf               -- } &{\bf                0.14 } \\[-4pt]
            4922.81 &               0.07 &{\bf               -- } &{\bf               -- } &{\bf               -- } &{\bf               -- } &{\bf               -- } \\[-4pt]
            4927.88 &{\bf                0.59 } &{\bf                0.40 } &{\bf                0.43 } &{\bf                0.42 } &{\bf                0.33 } &{\bf                0.35 } \\[-4pt]
            4934.20 &{\bf               -- } &{\bf               -- } &{\bf               -- } &{\bf                0.22 } &{\bf               -- } &{\bf               -- } \\[-4pt]
            4937.09 &               0.07 &{\bf               -- } &{\bf               -- } &{\bf               -- } &{\bf               -- } &{\bf               -- } \\[-4pt]
            4940.06 &{\bf                0.16 } &               0.10 &               0.10 &               0.09 &               0.04 &               0.06 \\[-4pt]
            4940.50 &               0.08 &{\bf               -- } &{\bf               -- } &{\bf               -- } &{\bf               -- } &{\bf               -- } \\[-4pt]
            4944.29 &{\bf                0.11 } &{\bf               -- } &{\bf               -- } &{\bf               -- } &{\bf               -- } &               0.03 \\[-4pt]
            4948.33 &               0.07 &{\bf               -- } &               0.09 &{\bf               -- } &{\bf               -- } &{\bf               -- } \\[-4pt]
            4954.60 &{\bf               -- } &{\bf               -- } &{\bf               -- } &{\bf               -- } &{\bf               -- } &{\bf                0.10 } \\[-4pt]
            4961.05 &{\bf                0.13 } &               0.10 &{\bf                0.15 } &               0.10 &               0.07 &               0.07 \\[-4pt]
            4964.14 &               0.08 &{\bf               -- } &{\bf               -- } &{\bf               -- } &{\bf               -- } &{\bf               -- } \\[-4pt]
            4966.29 &               0.09 &               0.09 &               0.11 &               0.08 &               0.08 &               0.04 \\[-4pt]
            4971.35 &{\bf                0.51 } &{\bf                0.29 } &{\bf                0.31 } &{\bf                0.32 } &{\bf                0.27 } &{\bf                0.30 } \\[-4pt]
            4990.45 &{\bf                0.27 } &{\bf                0.18 } &{\bf                0.22 } &{\bf                0.19 } &{\bf                0.16 } &{\bf                0.13 } \\[-4pt]
            5013.93 &{\bf                0.17 } &{\bf                0.13 } &               0.09 &               0.11 &               0.08 &               0.08 \\[-4pt]
            5016.89 &{\bf                0.44 } &{\bf                0.28 } &{\bf                0.27 } &{\bf                0.27 } &{\bf                0.22 } &{\bf                0.22 } \\[-4pt]
            5031.18 &               0.06 &{\bf               -- } &               0.07 &               0.06 &{\bf               -- } &{\bf               -- } \\[-4pt]
            5036.28 &{\bf                0.43 } &{\bf                0.30 } &{\bf                0.31 } &{\bf                0.29 } &{\bf                0.22 } &{\bf                0.22 } \\[-4pt]
            5041.34 &               0.07 &               0.09 &               0.07 &               0.06 &               0.05 &               0.07 \\[-4pt]
            5041.44 &               0.08 &               0.11 &               0.09 &               0.09 &               0.05 &               0.07 \\[-4pt]
            5088.16 &{\bf                0.18 } &               0.10 &               0.11 &               0.10 &               0.06 &{\bf                0.09 } \\[-4pt]
            5092.29 &               0.07 &{\bf               -- } &               0.04 &               0.04 &               0.04 &               0.05 \\[-4pt]
            5097.49 &{\bf                0.38 } &{\bf                0.24 } &{\bf                0.24 } &{\bf                0.23 } &{\bf                0.19 } &{\bf                0.18 } \\[-4pt]
            5136.27 &               0.08 &               0.05 &               0.08 &{\bf               -- } &{\bf               -- } &{\bf               -- } \\[-4pt]
            5140.82 &{\bf                0.14 } &               0.08 &               0.09 &               0.09 &               0.04 &               0.05 \\[-4pt]
            5156.66 &{\bf                0.13 } &{\bf               -- } &               0.06 &               0.06 &               0.05 &               0.07 \\[-4pt]
            5157.21 &               0.08 &{\bf               -- } &{\bf               -- } &{\bf               -- } &{\bf               -- } &{\bf               -- } \\[-4pt]
            5169.30 &{\bf                0.14 } &               0.08 &               0.08 &               0.08 &               0.07 &               0.06 \\[-4pt]
            5170.77 &{\bf                0.29 } &               0.10 &               0.10 &               0.10 &               0.09 &               0.08 \\[-4pt]
            5179.78 &               0.03 &{\bf               -- } &{\bf               -- } &{\bf               -- } &{\bf               -- } &{\bf                0.10 } \\[-4pt]
            5181.32 &{\bf                0.16 } &{\bf               -- } &{\bf               -- } &{\bf               -- } &{\bf               -- } &{\bf               -- } \\[-4pt]
            5212.22 &               0.08 &{\bf               -- } &               0.06 &{\bf               -- } &{\bf               -- } &{\bf               -- } \\[-4pt]
            5214.61 &{\bf                0.11 } &               0.05 &               0.03 &               0.04 &               0.05 &               0.03 \\[-4pt]
            5215.57 &{\bf                0.14 } &{\bf               -- } &{\bf               -- } &{\bf               -- } &{\bf               -- } &{\bf               -- } \\[-4pt]
            5217.89 &{\bf                0.13 } &{\bf               -- } &{\bf               -- } &{\bf               -- } &{\bf               -- } &{\bf               -- } \\[-4pt]
            5221.03 &{\bf                0.20 } &               0.12 &               0.12 &               0.09 &               0.09 &{\bf                0.10 } \\[-4pt]
            5221.75 &{\bf                0.12 } &               0.04 &{\bf               -- } &               0.06 &{\bf               -- } &               0.03 \\[-4pt]
            5225.81 &{\bf                0.14 } &{\bf               -- } &{\bf               -- } &{\bf               -- } &{\bf               -- } &{\bf               -- } \\[-4pt]
            5226.13 &               0.08 &               0.05 &               0.02 &               0.04 &               0.03 &{\bf               -- } \\[-4pt]
            5228.10 &{\bf                0.12 } &{\bf               -- } &{\bf               -- } &{\bf               -- } &{\bf               -- } &{\bf               -- } \\[-4pt]
            5242.06 &{\bf                0.15 } &               0.06 &{\bf               -- } &{\bf               -- } &{\bf               -- } &               0.06 \\[-4pt]
            5243.18 &{\bf                0.10 } &               0.06 &{\bf               -- } &{\bf               -- } &{\bf               -- } &               0.03 \\[-4pt]
            5248.98 &               0.08 &               0.09 &               0.10 &               0.08 &               0.07 &               0.05 \\[-4pt]
            5255.70 &               0.07 &               0.08 &               0.10 &               0.12 &               0.05 &               0.06 \\[-4pt]
            5263.77 &               0.07 &               0.09 &{\bf               -- } &{\bf               -- } &{\bf               -- } &{\bf               -- } \\[-4pt]
            5272.00 &               0.09 &{\bf               -- } &{\bf               -- } &{\bf               -- } &{\bf               -- } &{\bf               -- } \\[-4pt]
            5274.53 &{\bf                0.10 } &               0.06 &               0.05 &               0.06 &{\bf               -- } &               0.04 \\[-4pt]
            5275.13 &{\bf                0.22 } &{\bf               -- } &{\bf               -- } &{\bf               -- } &{\bf               -- } &{\bf               -- } \\[-4pt]
            5276.17 &               0.07 &{\bf               -- } &               0.08 &               0.08 &{\bf               -- } &               0.06 \\[-4pt]
            5278.78 &               0.08 &{\bf               -- } &{\bf               -- } &{\bf               -- } &{\bf               -- } &{\bf               -- } \\[-4pt]
            5281.32 &{\bf                0.10 } &{\bf               -- } &{\bf               -- } &{\bf               -- } &{\bf               -- } &{\bf               -- } \\[-4pt]
            5284.34 &{\bf               -- } &{\bf               -- } &{\bf                0.21 } &{\bf                0.16 } &{\bf               -- } &{\bf               -- } \\[-4pt]
            5298.51 &{\bf                0.11 } &{\bf               -- } &{\bf               -- } &{\bf               -- } &{\bf               -- } &{\bf               -- } \\[-4pt]
            5299.97 &{\bf                0.11 } &              -- &{\bf               -- } &{\bf               -- } &{\bf               -- } &{\bf               -- } \\[-4pt]
            5303.84 &               0.06 &{\bf               -- } &{\bf               -- } &{\bf               -- } &{\bf               -- } &{\bf               -- } \\[-4pt]
            5314.92 &               0.07 &{\bf               -- } &{\bf               -- } &               0.03 &{\bf               -- } &{\bf               -- } \\[-4pt]
            5341.15 &{\bf                0.27 } &{\bf                0.30 } &{\bf                0.25 } &{\bf                0.29 } &{\bf                0.23 } &{\bf                0.22 } \\[-4pt]
            5370.30 &               0.06 &               0.04 &               0.06 &{\bf               -- } &{\bf               -- } &               0.04 \\[-4pt]
            5389.85 &               0.09 &{\bf               -- } &               0.08 &{\bf               -- } &{\bf               -- } &{\bf               -- } \\[-4pt]
            5390.53 &{\bf                0.19 } &{\bf               -- } &               0.05 &               0.05 &               0.04 &               0.06 \\
\enddata
\tablecomments{Depth values larger than 3$\sigma$ are in bold}
\end{deluxetable*}

\begin{deluxetable}{lrrr}[ht!]
\tablecaption{Number of unidentified features \label{tab:nfeature}}
\tablecolumns{4}
\tablewidth{0pt}
\tablehead{
\colhead{Star} & \colhead{N$_{{\rm tot}}$} & \multicolumn{1}{c}{N$_{2\sigma-3\sigma}$} & \colhead{N$_{>3\sigma}$}  {} 
}
\startdata
 Sun &    67 & 25 &   35   \\
 00241708-7206106 &    36  & 11 &   10   \\
 00251219-7208053 &    42 & 12 &   12   \\
 00240054-7208550 &    40 & 11 &   12   \\
 02561410-0029286 &    30 & 8 &   9   \\
 03173493-0022132 &    39 & 10 &   17   \\
\enddata
\end{deluxetable}


\clearpage




\begin{thebibliography}{}
\bibitem[{Anders \& Grevesse (1989)}]{AND89} Anders,E. \& Grevesse, N. 1989, \gca, 53, 197	
\bibitem[{Brahm et al. (2017)}]{BRAHM17} Brahm, R., Jord\'an, A.; Hartman, J., \& Bakos, G. 2017, \mnras, 467, 971 (B17)
\bibitem[{Castelli (2005)}]{CA05} Castelli, F., 2005, Mem. Soc. Astron. Ital. Suppl., 8, 34
\bibitem[{Castelli \& Kurucz (2003)}]{CA03} Castelli, F.,  Kurucz, R. L. 2003, in IAU Symp. 210, Modelling of Stellar Atmosphere, ed. N. E. 
   Piskunov,W.W.Weiss, \& D. F. Gray (San Francisco, CA: ASP), A20 
\bibitem[{Coelho et al. (2005)}]{COELHO05} Coelho P., Barbuy B., Mel\'endez J., et al. 2005, \aap, 443, 735 (C05)
\bibitem[{Coelho (2014)}]{COELHO14} Coelho, P. R. T. 2014, \mnras, 440, 1027 (C14)
\bibitem[{Dekker et al. (2000)}]{DEK00} Dekker H., D’Odorico S., Kaufer A., Delabre B., Kotzlowski H., in Iye M., Moorwood A. F., eds,
 Proc. SPIE Conf. Ser. Vol. 4008, Optical and IR Telescope Instrumentation and Detectors. SPIE, Bellingham, p. 534
\bibitem[{de Laverny et al. (2012)}]{deLA12} de Laverny, P., Recio-Blanco, A., Worley, C. C., Plez, B. 2012, \aap, 544, A126 (AMBRE)
\bibitem[{De Silva et al. (2015)}]{deSI15} De Silva, G. et al. 2015, \mnras, 449, 2604
\bibitem[{Franchini et al. (2010)}]{FR10} Franchini M., Morossi C., Di Marcantonio P., Malagnini M. L., Chavez M., 2010, ApJ, 719, 240
\bibitem[{Garc\'ia P\'erez,  et al. (2015)}]{GA15} Garc\'ia P\'erez, A. E.,  et al. 2015, \aj, 151, 144
\bibitem[{Gilmore et al. (2012)}]{GIL12} Gilmore, G. et al. 2012, The Messenger, 147, 25
\bibitem[{Gonneau et al. (2016)}]{GO16} Gonneau, A. et al. 2016, \aap, 589, A36
\bibitem[{Gorgas et al. (1993)}]{GO93} Gorgas, J. et al. 1993, \apjs, 86, 153
\bibitem[{Gray \& Corbally (1994)}]{GR94} Gray, R. O., \& Corbally, D. J. L. I. 1994, \aj, 107, 742
\bibitem[{Grevesse et al. (2007)}]{GRE07} Grevesse, N., Asplund, M., Sauval, A. J. 2007, \ssr, 130, 105
\bibitem[{Gustafsson et al. (2008)}]{GU08} Gustafsson, B., Edvardsson, B., Eriksson, K., et al. 2008, \aap, 486, 951
\bibitem[{Hauschildt  \& Baron (1999)}]{HAU99} Hauschildt, P. H., \& Baron, E. 1999, J. Computat. Appl. Math., 109, 41 
\bibitem[{Hubeny (1988)}]{HUBENY88} Hubeny, I. 1988 Computer Physics Communications, Volume 52, Issue 1, p. 103-132
\bibitem[{Husser et al. (2013)}]{HUSSER13} Husser, T.-O., Wende-von Berg, S., Dreizler, S., et al. 2013, \aap 553, A6 (PHOENIX)
\bibitem[{Jacoby, Hunter, \& Christian (1984)}]{JACOBY84} Jacoby, G. H.; Hunter, D. A.; Christian, C. A. 1984, \apjs, 56, 257
\bibitem[{Jofr\'e et al. (2015)}]{JO15} Jofr\'e et al. 2015, \aap, 582, A81
\bibitem[{Kim et al. (2016)}]{KI16}Kim, H. et al. 2016, AJSS, 227, 24
\bibitem[{Korn, Maraston \& Thomas (2005)}]{KO05} Korn A.,  Maraston C., Thomas D. 2005, \aap, 438, 685
\bibitem[{Kos et al. (2017)}]{KO17} Kos, J., et al. 2017, \mnras, 464, 1259
\bibitem[{Kurucz (1979)}]{KURUCZ79} Kurucz, R. L. 1979, \apjs, 40, 1
\bibitem[{Kurucz (2005a)}]{KU05a} Kurucz, R. L. 2005a, Mem. Soc. Astron. Ital. Suppl., 8, 14
\bibitem[{Kurucz (2005b)}]{KU05b} Kurucz, R. L. 2005b, Mem. Soc. Astron. Ital. Suppl., 8, 76
\bibitem[{Kurucz (2014)}]{KUR14} Kurucz, R. L. 2014, in ``Determination of Atmospheric Parameters of B-, A-, F- and G-Type Stars.'' Series: 
GeoPlanet: Earth and Planetary Sciences, ISBN: 978-3-319-06955-5. Springer International Publishing (Cham), Edited by Ewa Niemczura, 
Barry Smalley and Wojtek Pych, pp. 63-73
\bibitem[{Leitherer et al. (1996)}]{LEITHERE96} Leitherer, C. Alloin, D., Alvensleben, U. F.-v,  et al.  1996, \pasp, 108, 996
\bibitem[{Lobel (2011)}]{LO11} Lobel, A.2011, CaJPh, 89,395
\bibitem[{Magrini et al. (2017)}]{MA17} Magrini, L., Randich, S., Kordopatis, G. et al. 2017, \aap, 603A, 2 
\bibitem[{Majewski et al. (2017)}]{MAJ17} Majewski, Steven R., Schiavon, Ricardo P., Frinchaboy, Peter M., el al. 2017, \aj, 154, 94
\bibitem[{M\'{e}sz\'{a}ros \& Allende Prieto (2013)}]{MES13} M\'{e}sz\'{a}ros, Sz. \& Allende Prieto, C. 2013, \mnras, 430, 3285
\bibitem[{Molaro et al. (2013)}]{MO13} Molaro, P., Monaco, L., Barbieri, M., Zaggia, S. 2013, \mnras, 429, 79
\bibitem[{Moultaka et al. (2004)}]{MO04} Moultaka, J., Ilovaisky, S. A., Prugniel, P.,  Soubiran, C. 2004, \pasp, 116, 693 (ELODIE)
\bibitem[{Munari et al. (2005)}]{MU05} Munari, U., Sordo, R., Castelli, F., \& Zwitter, T.  2005, \aap, 442, 1127
\bibitem[{Partridge \& Schwenke (1997)}]{PA97} Partridge, H., \&  Schwenke, D. W. 1997, \jcp, 106, 4618
\bibitem[{Prugniel \& Soubiran (2001)}]{PRU01} Prougniel, Ph. \& Soubiran, C. 2001, \aap, 369, 1048
\bibitem[{Prugniel et al. (2007)}]{PRU07} Prugniel,Ph et al. 2007, arXiv:astro-ph/0703658
\bibitem[{Randich \& Gilmore (2013)}]{RAN13} Randich, S., Gilmore, G., Gaia-ESO Consortium 2013,  The Messenger, 154,47
\bibitem[{Sacco et al. (2014)}]{SA14} Sacco G. G. et al. 2014, \aap, 565,513 
\bibitem[{S\'anchez-Bl\'azquez et al. (2006)}]{SA06} S\'anchez-Bl\'azquez et al. 2006, \mnras, 371, 703 (MILES)
\bibitem[{Schwenke (1998)}]{SC98} Schwenke, D. W. 1998, Faraday Discuss., 109, 321
\bibitem[{Smiljanic et al. (2014)}]{SM14} Smiljanic R. et al. 2014, \aap, 570, A122
\bibitem[{Sousa et al. (2015)}]{SO15} Sousa, S. G. et al. 20145, \aap, 577, A67 
\bibitem[{Swan (1957)}]{SW57} Swan, W. 1857, Transactions of the Royal Society of Edinburgh, 21, 411
\bibitem[{Thomas,  Maraston \& Korn (2004)}]{TH04} Thomas D., Maraston C., Korn A. 2004, \mnras, 351, L19
\bibitem[{Valdes et al. (2004)}]{VA04} Valdes, F. et al. 2004, \apjs, 152, 251 (INDO-U.S.)
\bibitem[{Vazdekis et al. (2015)}]{VA15} Vazdekis, A. et al. 2004, \mnras, 449, 1177 (INDO-U.S.)
\bibitem[{Weck et al. (2003)}]{WE03} Weck, P. F. et al. 2003, \apj, 582, 1059 
\bibitem[{Worley et al. (2016)}]{WO16} Worley, C. C. et al. 2016, \aap, 591, A81
\bibitem[{Worthey et al. (1994)}]{WR94} Worthey, G. et al. 1994, \apjs, 94, 687
\bibitem[{Worthey \& Ottaviani (1997)}]{WR97} Worthey G. \& Ottaviani D. L., 1997, ApJS, 111, 377
\bibitem[{Worthey, Danilet \& Faber (2014)}]{WR14} Worthey, G., Danilet, A. B., Faber, S. M., 2014, \aap 561, A36
\bibitem[{Yanny et al. (2009)}]{YANNY09} Yanny, B., Rockosi, C., Newberg, H. J., et al. 2009, \aj, 137, 4377
\bibitem[{York et al. (2000)}]{YOR00} York, D. G. et al. 2000,\aj, 120, 1579

\end{thebibliography}
\end{document}